\documentclass[%
aip,
amsmath,amssymb,
groupedaddress,
 reprint,%
]{revtex4-1}

\usepackage{graphicx}
\usepackage{dcolumn}
\usepackage{bm}
\usepackage{color}
\usepackage[utf8]{inputenc}
\usepackage[T1]{fontenc}
\usepackage{mathptmx}
\DeclareSymbolFont{newfont}{OML}{cmm}{m}{it}
\DeclareMathSymbol{\varrho}{3}{newfont}{37}
\usepackage{etoolbox}
\usepackage{empheq}

\renewcommand{\d}{{\rm d}}

\newcommand{\w}{\omega}

\newcommand{\ti}{\tilde}
\newcommand{\B}{\mbox{\tiny B}}

\newcommand{\tB}{\mbox{\tiny B}}
\newcommand{\tP}{{\rm core}}

\newcommand{\tS}{\mbox{\tiny S}}
\newcommand{\T}{\mbox{\tiny T}}

\newcommand{\SB}{\mbox{\tiny SB}}
\newcommand{\dg}{\dagger}
\newcommand{\la}{\langle}
\newcommand{\ra}{\rangle}

\newcommand{\La}{\big\la}
\newcommand{\Ra}{\big\ra}

\newcommand{\tFP}{\mbox{\tiny FP}}

\newcommand{\eg}{\mbox{\tiny eg}}

\newcommand{\Sec}[1]{Sec.\,\ref{#1}}

\newcommand{\nl}{\nonumber \\}
\newcommand{\be}{\begin{equation}}
\newcommand{\ee}{\end{equation}}
\newcommand{\bsube}{\begin{subequations}}
\newcommand{\esube}{\end{subequations}}
\newcommand{\Eq}[1]{Eq.\,(\ref{#1})}
\newcommand{\Eqs}[1]{Eqs.\,(\ref{#1})}
\newcommand{\Fig}[1]{Fig.\,\ref{#1}}

\newcommand{\RN}[1]{%
  \textup{\uppercase\expandafter{\romannumeral#1}}%
}

\newcommand{\greater}{\mbox{\tiny $ >$}}
\newcommand{\lesser}{\mbox{\tiny $ <$}}
\newcommand{\plus}{\mbox{\tiny $+$}}
\newcommand{\minus}{\mbox{\tiny $-$}}
\newcommand{\lgter}{\mbox{\tiny $\pmb \lessgtr$}}
\newcommand{\gler}{\mbox{\tiny $\pmb \gtrless$}}

\begin{document}

\title{
	%
	Open quantum systems with nonlinear environmental backactions: Extended dissipaton theory versus core--system hierarchy construction
}
\author{Zi-Hao Chen}
\author{Yao Wang}     \email{wy2010@ustc.edu.cn}
\author{Rui-Xue Xu}   \email{rxxu@ustc.edu.cn}
\author{YiJing Yan}
\affiliation{
	Hefei National Research Center for Physical Sciences at the Microscale and Department of Chemical Physics,
	University of Science and Technology of China, Hefei, Anhui 230026, China
}
\date{25 January 2023}

\begin{abstract}
	In this paper, we present a comprehensive account of quantum dissipation theories with the quadratic environment couplings.
	The theoretical development includes the Brownian solvation mode embedded hierarchical quantum master equations,  a core--system hierarchy construction that verifies the extended dissipaton equation of motion (DEOM) formalism [R.\ X.\ Xu \emph{et al}., J.\ Chem.\ Phys.\ {\textbf{148}}, 114103 (2018)].
	Developed are also the quadratic imaginary--time DEOM for equilibrium and the $\lambda(t)$-DEOM for nonequilibrium thermodynamics problems.
	Both the celebrated Jarzynski equality and Crooks relation are accurately reproduced, which in turn confirms the rigorousness of the extended DEOM theories.
	While the extended DEOM is more numerically efficient, the core--system hierarchy quantum master equation is favorable for ``visualizing'' the correlated solvation dynamics.
\end{abstract}
\maketitle

\section{Introduction}
\label{sec1}
Quantum dissipation plays a crucial role in many fields of modern science, where irreversibility takes place during relaxation, dephasing, transport and thermodynamic processes.\cite{Wei21, Kle09, Bre02, Yan05187, Nit06, Muk95, Lou73, Haa7398, Aka15056002, Che964565,Imr02, Hau08}
In these studies, the environmental non-Markovian and nonperturbative quantum nature would be prominent if the system and bath are strongly correlated.
Various exact methods, such as the Feynman--Vernon influence functional approach \cite{Fey63118} and its differential equivalence, the {hierarchical equations of motion} (HEOM) formalism,\cite{Tan906676, Tan06082001, Yan04216, Xu05041103, Xu07031107, Jin08234703, Ye16608, Yan16110306,Yan20204109} had been constructed.
There are also quite a few studies that treat HEOM as quantum Fokker--Planck (FP) type equations, via transferring the involving degrees of freedom into the
Wigner representation, such as the quantum hierarchical FP equation.\cite{Tan914131,Tan943049,
Tan971779,Ike17014102,Ike192517,Tan15144110}
Besides,  transformations from HEOM to the low--temperature quantum FP or  Smoluchowski equations are also suggested.\cite{Li22064107,Ike22104104}
However, most of these theories are exact only for Gaussian environments with linear couplings.
This linearity intrinsically implies a weak backaction of the central system on the surroundings.

On the other hand, the nonlinear system--bath interactions are generally common and appealing in real physical systems.\cite{Vio02886, Mak04178301, Mul04237401, Ber05257002, Yan865908, Pen07114302, Wan0710369, Zha121075, Cho17074114, Wan21462}
The quest of an exact treatment of quantum dissipation with nonlinear environment couplings remains a challenging task in recent years.\cite{Xu17395, Xu18114103, Yan19074106, Hsi18014104, Hsi20125002, Hsi20125003}
In this account, we consider the total system--plus--bath composite Hamiltonian to take the form of
\be\label{HT0}
H_{{\T}}=H_{\tS} + h_{\B}
+ \hat Q_{\tS}(\alpha_0+\alpha_1\hat x_{\B}+ \alpha_2\hat x_{\B}^{2}).
\ee
Here, $H_{\tS}$ is the system Hamiltonian and $\hat Q_{\tS}$ is the dissipative mode which can be an arbitrary Hermitian system operator.
The bath Hamiltonian and solvation coordinate are
\be\label{hB0}
h_{\B}= \frac{1}{2}\sum_j \omega_j(\hat p_j^2 + \hat q_j^2)
\ \ \text{and}\ \
\hat x_{\B}=\sum_j c_j \hat q_j,
\ee
respectively,  
where the Gaussian $h_{\B}$ is in line with the central limiting statistics description.
The  solvation coordinate $\hat x_{\B}$ is defined to be the linear part, but involved in both the $\alpha_1$ and $\alpha_2$ terms.
In \Eq{HT0},  $\hat Q_{\tS}$ and $\hat x_{\B}$ are set to be dimensionless, while the $\alpha$--parameters are of energy unit.
When $\alpha_2=0$, it is reduced to the linear bath coupling case.
The nonlinearity is exemplified here with the quadratic coupling.
The approaches presented later in this account can all be extended to higher orders.
Throughout this paper we set $\hbar=1$ and $\beta=1/(k_BT)$, with $k_B$ being the Boltzmann constant and $T$ the temperature.

To simulate the dynamics governed by the Hamiltonian in \Eqs{HT0} with (\ref{hB0}), we had previously adopted the dissipaton theory.
The dissipaton theory introduces statistical quasi-particles, dissipatons, to characterize the interacting bath statistical properties. \cite{Yan14054105, Xu151816, Zha18780, Wan20041102, Wan22170901} The resulting dissipaton equation of motion (DEOM) is not only identical to the HEOM for the reduced system dynamics, but also convenient to treat the hybridized bath dynamics.\cite{Zha15024112, Che21244105, Wan22170901}
Based on the DEOM, we had proposed two distinct approaches for the nonlinear bath coupling, namely,
the extended DEOM \cite{Xu17395, Xu18114103} and the stochastic--fields--dressed DEOM (SFD--DEOM). \cite{Che21174111}
%
%

In this work, an exact core--system hierarchy construction is developed. It explicitly treats the solvation phase space, including the nonlinear coupling term ($\alpha_2$ term).
We name it as  
the Brownian solvation mode embedded hierarchical quantum master equations (BSM-HQME), with the standard FP algebra\cite{Ris89,Xu09074107,Din17024104}
 being exploited.
%
%
We further scrutinize the extended DEOM \cite{Xu17395, Xu18114103} and the SFD--DEOM,\cite{Che21174111} with the newly developed BSM-HQME.
All of them agree with each other,  
as inferred from their theoretical constructions and also  evident from numerical simulations.
This implies the extended DEOM and the underlying generalized Wick's theorem (GWT-2) \cite{Xu17395, Xu18114103} are universally correct.

Some features of these methods are as follows.
(\emph{i}) The SFD-DEOM is exact in principle, constructed on the basis of SFD total Hamiltonian with only linear bath coupling terms, while the nonlinear terms are resolved via the stochastic fields. However, the SFD-DEOM approach is numerically available only for short--time evolutions, but subject to long--time instability.
(\emph{ii}) The BSM-HQME explicitly treats the solvation coordinate and momentum. Its phase space dynamics can be directly computed using this approach.
Taking the electron transfer process, for example, this approach provides the exact reaction coordinate evolutions.
(\emph{iii}) The extended DEOM is mostly numerically efficient among those three methods. It can be readily extended to imaginary--time ($i$-DEOM) formulations \cite{Gon20214115, Wan22170901,Tan14044114} 
to compute the hybridization free energy.
(\emph{iv})
 We can also develop the nonequilibrium $\lambda$-DEOM (neq-$\lambda$-DEOM) \cite{Gon22054109, Wan22170901} to investigate the fluctuation theorems, such as Jarzynski equality and Crooks relation.
The results here in turn confirm the rigorousness of the extended DEOM theories.

The remainder of this paper is organized as follows.
Section \ref{sec_eDEOM} comprises a complete description of extended DEOM.
In \Sec{sec_fp_a}, the BSM-HQME is developed in detail.
The FP algebra is outlined in Appendix \ref{FP}.
The numerical cross-check is carried out among BSM-HQME, extended DEOM, and SFD-DEOM in terms of both real--dynamics and spectroscopies in \Sec{sec_fp_b}.
The explicit solvation mode dynamics are also computed using the BSM-HQME.
Section \ref{sec_therm} is concerned with the system--bath thermodynamic mixing with nonlinear environment couplings.
The Jarzynski equality and Crooks relation are accurately reproduced with extended DEOM for quadratic bath couplings.
We summarize this work in \Sec{sec_sum}.

\section{The extended DEOM formalism}\label{sec_eDEOM}
In this section, we briefly review the extended DEOM formalism in Refs.\,\onlinecite{Xu17395, Xu18114103} for nonlinear bath couplings. Before that, we first introduce the associated bath statistics.
With the bath Hamiltonian and coupling in the form of \Eqs{HT0} and (\ref{hB0}), the bath influence is completely described via the bath spectral density,
\be\label{chiJw}
J(\w\geq 0) = \frac{\pi}{2}\sum_j c^2_j\delta(\w-\w_j) = -J(-\w).
\ee
It is expressed in terms of $J(\w)\equiv {\rm Im}\chi_{\B}(\w)$, with
\cite{Yan05187, Wei21}
\begin{align}\label{chiw_def}
	\chi_{\B}(\w) & =i\int^{\infty}_{0}\!{\rm d}t\,e^{i\w t}
	\La [\hat x_{\B}(t),\hat x_{\B}(0)]\Ra_{\B}.
\end{align}
Here, $[\,\cdot\,,\cdot\,]$ denotes a commutator, $\hat x_{\B}(t)\equiv e^{ih_{\B}t}\hat x_{\B}e^{-ih_{\B}t}$ and $\la \hat O \ra_{\B} \equiv {\rm tr}_{\B}(\hat O e^{-\beta h_{\B}})/{\rm tr}_{\B}e^{-\beta h_{\B}}$.
The fluctuation--dissipation theorem in relation to \Eq{chiw_def} reads \cite{Wei21, Yan05187}
\be\label{FDT}
\la \hat x_{\B}(t)\hat x_{\B}(0)\ra_{\B}
= \frac{1}{\pi} \int^{\infty}_{-\infty}\!\! {\rm d}\omega\,
\frac{e^{-i\omega t} J(\w)}{1-e^{-\beta\omega}}.
\ee
This bare--bath subspace relation holds in general, regardless of the nature of bare bath Hamiltonian and also independent of the system--bath couplings involved in \Eq{HT0}.

The DEOM formalism is established by expressing the influence of environment with a finite number of statistically independent quasi--particles, the dissipatons.\cite{Yan14054105,Yan16110306}
We expand \Eq{FDT} in an exponential series,\cite{Yan05187,Hu10101106,Hu11244106}
\be\label{FF_exp_all}
\la\hat x_{\B}(t)\hat x_{\B}(0)\ra_{\B} = \sum_{k=1}^K \eta_k e^{-\gamma_k t}.
\ee
Its time reversal is expressed in the form of\cite{Yan05187, Din12224103}
\be\label{time-reversal}
\la\hat x_{\B}(0)\hat x_{\B}(t)\ra_{\B} = \la\hat x_{\B}(t)\hat x_{\B}(0)\ra^{\ast}_{\B}  = \sum_{k=1}^K \eta^{\ast}_{\bar k}e^{-\gamma_k t},
\ee
with $\bar k$ being defined via $\gamma_{\bar k}\equiv \gamma^{\ast}_{k}$, which must also appear in \Eq{FF_exp_all}.
The solvation coordinate can then be recast in the dissipatons decomposition form,\cite{Yan14054105, Yan16110306}
\be\label{xB_in_f}
\hat x_{\B} = \sum_{k=1}^K \hat f_{k},
\ee
with
\bsube\label{ff_t}
\begin{align}\label{ff_t_a}
	\la \hat f_{k}(t)\hat f_{k'}(0)\ra_{\B} & = \la \hat f_{k}\hat f_{k'}\ra^{>}_{\B} e^{-\gamma_k t} = \delta_{kk'}\eta_{k}e^{-\gamma_k t},
	\\ \label{ff_t_b}
	\la \hat f_{k'}(0)\hat f_{k}(t)\ra_{\B} & = \la \hat f_{k'}\hat f_{k}\ra^{<}_{\B} e^{-\gamma_k t} = \delta_{kk'}\eta^{\ast}_{\bar k}e^{-\gamma_k t}.
\end{align}
\esube
Apparently, both \Eqs{FF_exp_all} and (\ref{time-reversal}) are reproduced.

Dynamical variables in DEOM are the dissipaton density operators (DDOs):\cite{Yan14054105, Yan16110306}
\be\label{rhon}
\rho^{(n)}_{\bf n}(t)\equiv \rho^{(n)}_{n_1\cdots n_K}(t)\equiv
{\rm tr}_{\B}\big[
	(\hat f_{K}^{n_K}\cdots\hat f_{1}^{n_1})^{\circ}\rho_{\T}(t)\big].
\ee
The reduced system density operator is just $\rho_{\tS}(t)\equiv \rho^{(0)}_{\bf 0}(t)$.
The indexes ${\bf n}\equiv \{n_1\cdots n_K\}$ and $n=n_1+\cdots +n_K$ specify the occupations and the total number of dissipatons, respectively.
The notation, $(\cdots)^{\circ}$, denotes the irreducible representation. We have $(\hat f_k\hat f_{k'})^{\circ}=(\hat f_{k'} \hat f_k)^{\circ}$ for bosonic dissipatons.

The construction of DEOM starts from
\be\label{dot_rhon}
\dot\rho^{(n)}_{\bf n}(t)\equiv \dot\rho^{(n)}_{n_1\cdots n_K}(t)\equiv
{\rm tr}_{\B}\big[
	(\hat f_{K}^{n_K}\cdots\hat f_{1}^{n_1})^{\circ}
		{\dot\rho}_{\T}(t)\big],
\ee
with the total composite density operator satisfying
\be\label{SchEq}
\dot\rho_{\T}(t) = -i[H_{\T}, \rho_{\T}(t)].
\ee
The dissipaton formalism consists of the generalized diffusion equation and the generalized Wick's theorems (GWTs). The former reads
\be\label{diff_0}
{\rm tr}_{\B}\Big[\Big(\frac{\partial \hat f_k}{\partial t}\Big)_{\B}\rho_{\T}(t)\Big] = -\gamma_{k}\,{\rm tr}_{\B}\big[\hat f_k\rho_{\T}(t)\big].
\ee
It together with $\big(\frac{\partial \hat f_k}{\partial t}\big)_{\B}=-i[\hat f_k, h_{\B}]$ will give rise to
\be\label{diff}
i\,{\rm tr}_{\B}\big\{
(\hat f_{K}^{n_K}\cdots\hat f_{1}^{n_1})^{\circ} [h_{\B},\rho_{\T}]\big\}
= \Big(\sum_{k=1}^{K} n_k\gamma_k\Big)\rho^{(n)}_{\bf n}.
\ee

The GWT-1 evaluates the linear bath coupling with one dissipaton added each time. It reads\cite{Yan14054105,Yan16110306}
\begin{align}\label{Wick10} 
	    & \quad\,{\rm tr}_{\B}\big[(\hat f_{K}^{n_K}\cdots\hat f_{1}^{n_1})^{\circ}
		\hat f_\kappa\rho_{\T}(t)\big]
	\nl & = \rho^{(n+1)}_{{\bf n}^{+}_{\kappa}}(t)
	+ \sum_{k=1}^{K}
	n_k \la \hat f_{k}\hat f_{\kappa}\ra^{>}_{\B}\rho^{(n-1)}_{{\bf n}^{-}_{k}}(t).
\end{align}
The expression of ${\rm tr}_{\B}\big[(\hat f_{K}^{n_K}\cdots\hat f_{1}^{n_1})^{\circ}
		\rho_{\T}(t)\hat f_\kappa\big]$
is similar, but with  $\la \hat f_{k}\hat f_{\kappa}\ra^{>}_{\B}$ being replaced by $\la \hat f_{\kappa}\hat f_{k}\ra^{<}_{\B}$.
The associated index ${\bf n}^{\pm}_{k}$ differs from ${\bf n}\equiv \{n_1\cdots n_K\}$ by replacing the specified $n_k$ with $n_k\pm 1$.
This specifies the $(n\pm 1)$--particle DDO, $\rho^{(n \pm 1)}_{{\bf n}^{\pm}_{k}}(t)$, in \Eq{Wick10}.
In comparison, we may recall some properties about the
``normal order'' in textbooks,
which arranges creation operators
before annihilation operators.
Denote this also with $(\,\cdot\,)^{\circ}$,
such that $(\hat a^{\dg} \hat a )^{\circ}=(\hat a\hat a^{\dg})^{\circ}
=\hat a^{\dg} \hat a$.
Assume $\hat f= \sqrt{\eta}\,(\hat a+\hat a^{\dg})$,
with $\eta$ being an arbitrary real parameter.
It is easy to obtain
$(\hat f^n)^{\circ}\hat f=(\hat f^{n+1})^{\circ}+n\eta(\hat f^{n-1})^{\circ}
=\hat f(\hat f^n)^{\circ}$, in line with the standard normal ordering.
\cite{Fan112145}
The GWT-1 is just
the generalization of this result, which  has been already  verified analytically in the linear environmental coupling scenarios.\cite{Wan22170901,Wan20}

The GWT-2 is related to the quadratic bath coupling, where a pair of dissipatons are added each time.
It was  validated under the minimum--dissipaton ansatz,\cite{Din16204110,Din17024104} via the Zusman or the Fokker--Planck algebra.\cite{Xu17395}
One of the purposes of this work is to confirm the GWT-2 beyond these limitations. 
More precisely,  
we will first assume the correctness of the GWT-2 in more general scenarios, and  then   scrutinize it numerically  by comparing with the exact results from the  SFD--DEOM 
 \cite{Che21174111} and  the newly developed BSM-HQME (cf.\,\Sec{sec_fp}). In this sense, the GWT-2 is a  validated ``theorem'' that is generally correct. However, a rigorous proof of GWT-2 within the 
canonical  Feynman--Vernon influence functional formalism is absent so far.
%
%
The GWT-2 is evaluated as \cite{Xu17395, Xu18114103}
\begin{align}\label{Wick2}
	    & \quad\,{\rm tr}_{\B}\big[(\hat f_{K}^{n_K}\cdots\hat f_{1}^{n_1})^{\circ}
		(\hat f_\kappa\hat f_{\kappa'})\rho_{\T}(t)\big]
	\nl & = \rho^{(n+2)}_{{\bf n}^{++}_{\kappa\kappa'}}(t)
	+ \la\hat f_\kappa \hat f_{\kappa'}\ra_{\B}\rho^{(n)}_{{\bf n}}(t)
	+\sum_{k} n_k \la\hat f_k\hat f_{\kappa'}\ra^{>}_{\B}\rho^{(n)}_{{\bf n}^{-+}_{k\kappa}}(t)
	\nl & \quad
	+\sum_{k} n_k \la\hat f_k\hat f_\kappa\ra^{>}_{\B}\rho^{(n)}_{{\bf n}^{-+}_{k\kappa'}}(t)
	\nl & \quad
	+ \sum_{k,k'} n_k(n_{k'}-\delta_{kk'})\la\hat f_k\hat f_\kappa\ra^{>}_{\B}\la\hat f_{k'}\hat f_{\kappa'}\ra^{>}_{\B}
	\rho^{(n-2)}_{{\bf n}^{--}_{k k'}}(t).
\end{align}
The associated DDO index, ${\bf n}^{\pm\pm}_{k\kappa}$, differs from ${\bf n}\equiv n_1\cdots n_K$ on the specified subindexes, $n_k$ and $n_\kappa$, that are replaced by $n_k\pm 1$ and $n_\kappa\pm 1$, respectively.
Together with \Eqs{xB_in_f} and (\ref{ff_t}), we obtain
\begin{align}\label{Wick_xB2}
	    & \quad\,{\rm tr}_{\B}\big[(\hat f_{K}^{n_K}\cdots\hat f_{1}^{n_1})^{\circ}
	\hat x^2_{\B}\rho_{\T}(t)\big]
	\nl & = \sum_{kk'}\rho^{(n+2)}_{{\bf n}^{++}_{kk'}}(t)
	+ \la\hat x^2_{\B}\ra_{\B}\rho^{(n)}_{\bf n}(t)
	+ 2\sum_{kk'} n_k\eta_k\rho^{(n)}_{{\bf n}^{-+}_{kk'}}(t)
	\nl & \quad
	+ \sum_{kk'} n_k(n_{k'}-\delta_{kk'})\eta_k\eta_{k'}\rho^{(n-2)}_{{\bf n}^{--}_{kk'}}(t).
\end{align}
The expression of ${\rm tr}_{\B}\big[(\hat f_{K}^{n_K}\cdots\hat f_{1}^{n_1})^{\circ} \rho_{\T}(t)\hat x^2_{\B}\big]$ is similar, but with $\eta_k$ and $\eta_{k'}$ being replaced by $\eta^{\ast}_{\bar k}$ and $\eta^{\ast}_{\bar k'}$, respectively.

Combining \Eqs{dot_rhon}--(\ref{Wick_xB2}), the extended DEOM is finally constructed \cite{Xu17395, Xu18114103}
\begin{align} \label{gen_DEOM}
	\dot\rho^{(n)}_{\bf n} \!
	= & - \Big(i{\cal L}_{\tS} + \sum_{k}n_k\gamma_k\Big)\rho^{(n)}_{\bf n} - i\Big(\alpha_0+\alpha_2\la\hat x^2_{\B}\ra_{\B}\Big) {\cal A} \rho^{(n)}_{\bf n} \nl
	  & -i\alpha_1\sum_{k}\Big({\cal A}\rho^{(n+1)}_{{\bf n}^{+}_{k}} +n_k{\cal C}_k\rho^{(n-1)}_{{\bf n}^{-}_{k}}\Big)  -2 i \alpha_2 \sum_{kk'} n_k {\cal C}_k \rho^{(n)}_{{\bf n}^{-+}_{kk'}}\nl
	  & -i\alpha_2\sum_{kk'} \Big[{\cal A}\rho^{(n+2)}_{{\bf n}^{++}_{kk'}} + n_k(n_{k'}-\delta_{kk'}){\cal B}_{kk'}\rho^{(n-2)}_{{\bf n}^{--}_{kk'}} \Big],
\end{align}
with ${\cal L}_{\tS} \hat O \equiv \big[H_{\tS}, \hat O]$ and other involved superoperators defined as
\bsube\label{calABC}
\begin{align} \label{def_ABC}
	 & {\cal A} \hat O \equiv \big[\hat Q_{\tS}, \hat O],
	\\
	 & {\cal B}_{kk'} \hat O \equiv  \eta_k \eta_k' \hat Q_{\tS} \hat O
	- \eta^{\ast}_{\bar k} \eta^{\ast}_{\bar k'}\hat O \hat Q_{\tS},
	\\
	 & {\cal C}_k\hat O \equiv \eta_k \hat Q_{\tS}\hat O
	-\eta^{\ast}_{\bar k} \hat O \hat Q_{\tS}.
\end{align}
\esube

\section{BSM-HQME: A core-system hierarchy construction}\label{sec_fp}
In this section, we propose a new exact core--system hierarchy method, named as BSM-HQME.
In the core--system description, the bath Hamiltonian of \Eq{hB0} is divided into two parts, the solvation mode and the secondary bath coupled to it.
The FP algebra is used to provide a basis set for describing the solvation mode inside the core system. 
\cite{Ris89,Xu09074107,Din17024104}
The details of BSM-HQME construction are given in  \Sec{sec_fp_a}.
In \Sec{sec_fp_b}, we carry out the numerical cross check among BSM-HQME, extended DEOM, and SFD-DEOM. Both dynamics and spectroscopies  are simulated, together with the explicit solvation mode dynamics obtained from the BSM-HQME.

\subsection{The construction of BSM-HQME}\label{sec_fp_a}
In the core--system description, the bath Hamiltonian of \Eq{hB0} is divided into the solvation modes and secondary bath parts with the Caldeira--Leggett model, which reads
\be \label{CL}
h_{\B}=\frac{1}{2}\w_{\B}(\hat p_{\B}^2 + \hat x_{\B}^2) + \frac{1}{2}\sum_j \ti \w_{j}\big[\ti p_{j}^2+(\ti x_{j}-\frac{\ti c_{j}}{\ti \w_{j}}\hat x_{\B})^2\big].
\ee
The solvation mode behaves as a Brownian oscillator, with the correlation function, $\la \hat x_{\B}(t)\hat x_{\B}(0) \ra_{\B}$,
given by \Eq{FDT} in which\cite{Xu09074107,Din17024104}
\be\label{JBO}
J(\w)={\rm Im}\frac{\w_{\B}}{{\w}_{\B}^2-\w^2-i\w\zeta_{\B}(\w)}.
\ee
Here, $\zeta_{\B}(\w)\equiv \int_{0}^{\infty}\!{\rm d}t\,e^{i\w t}\ti\zeta_{\B}(t)$, with the classical friction function reading
\be \label{eta_2}
\ti\zeta_{\B}(t)=\w_{\B}\sum_{j}(\ti c_{j}^2/\ti \w_{ j})\cos(\ti \w_{j}t).
\ee
The spectral density of the secondary bath, $\ti h_{\B}=\frac{1}{2}\sum_j \ti \w_{j}\big(\ti p_{j}^2+\ti x_{j}^2\big)$, is then\cite{Xu09074107,Din17024104}
\be \label{spec2nd}
\ti J(\w)=\frac{\w}{\w_{\B}}{\rm Re}\zeta_{\B}(\w),
\ee
and the secondary bath fluctuation--dissipation theorem reads
\be \label{FD2nd}
\la \ti F(t)\ti F \ra_{\ti \B}=\frac{1}{\pi}
\int^{\infty}_{-\infty}\!\!{\rm d}\omega\,
\frac{\ti J(\w) e^{-i\omega t}}{1-e^{-\beta\omega}}
\ee
where $\la \hat O \ra_{\ti\B}\equiv {\rm tr}_{\ti\B} (\hat O e^{-\beta h_{\ti\B}})/{\rm tr}_{\ti\B} e^{-\beta h_{\ti\B}}$.
%
%

Using \Eq{FD2nd} with Eq.(\ref{spec2nd}), we can write down the HEOM/DEOM where the primary system and solvation mode compose the core system,
\begin{equation} \label{core_system}
	H_{\tP} = H_{\tS}+\hat Q_{\tS} F(\hat x_{\B}) + \frac{\w_{\B}}{2} (\hat p_{\B}^2 + \hat x_{\B}^2) + \ti \lambda {\hat x_{\B}}^2
\end{equation}
where
\be
F(\hat x_{\B})=\alpha_{0}+\alpha_1\hat x_{\B}+ \alpha_2\hat x_{\B}^{2}
\ee
and
\be \label{ti_lambda}
\ti \lambda\equiv \frac{1}{2} \sum_j \frac{{\ti c_j}^2}{{\ti \w}_j} = \frac{\ti\zeta_{\B}(0)}{2\w_{\B}}.
\ee
The secondary bath is treated as the environment. According to \Eq{FD2nd}, followed by applying the exponential decomposition scheme, \cite{Hu10101106, Hu11244106,Che22221102} we obtain
\be
\la \ti F(t)\ti F (0) \ra_{\ti \B} =\sum_{k=1}^{K}\ti\eta_{k}e^{-\ti\gamma_kt}\ \ \ \ (t>0)
\ee

The HEOM for the core system can be then constructed as
\begin{align}\label{DEOM}
	\dot{\rho}^{}_{\ti{\bf n}}=
	    & - (i{\cal L}_{\tP} +\gamma_{\,\ti{\bf n}})\rho^{}_{\ti{\bf n}} -i\sum_{k=1}^{K}\hat x_{\B}^{\times}\rho^{}_{{\ti{\bf n}}_{k}^+}
	\nl & -i\sum_{k=1}^{K}\ti n_{k} \big(\ti\eta_{k}\hat x^{\greater}_{\B}-\ti\eta^{\ast}_{\bar k} \hat x_{\B}^{\lesser}\big)\rho^{}_{\ti{\bf n}_{k}^-},
\end{align}
where ${\cal L}_{\tP} \hat O \equiv \big[H_{\tP}, \hat O]$ and $\gamma_{\,\ti{\bf n}}=\sum_{k}\ti n_k\ti\gamma_k$. The $\ti {\bf n}$ is the array of dissipaton occupation numbers; see comments after \Eq{rhon}.
Hereafter, we denote $A^{\times} \equiv A^{>} - A^{<}$, $A^{>} \hat O \equiv A \hat O$, and $A^{<} \hat O \equiv \hat O A$.
In \Eq{DEOM}, $\{\rho^{}_{\ti{\bf n}}\}$ are the DDOs of the core system, i.e.\ the primary system plus the solvation mode degrees of freedom.
Let $\hat W_{\ti{\bf n}}(x_{\B},p_{\B}) \equiv (\rho_{{\ti{\bf n}}})_{\mathrm{Wigner}}$ be the Wigner representation of the solvation subspace.
Moreover, we expand $\{\hat W_{\ti{\bf n}}(x_{\B},p_{\B})\}$ by the FP basis set as [cf.\ \Eq{doublerho}]
\begin{align}\label{expansion}
	\hat W_{\ti{\bf n}}(x_{\B},p_{\B}; t) = & \Big(\frac{\beta\w_{\B}}{2\pi}\Big)^{\frac{1}{2}} \sum_{n_{1},n_{2}} \frac{s_{n_{1},n_{2}}}{\sqrt{n_1!n_2!}} \rho_{n_{1},n_{2};\ti{\bf  n}}(t)
	\nl                                    &
	\times  e^{-\frac{\beta\w_{\B}}{4}(x_{\B}^2+p_{\B}^2)} \Psi_{n_{1},n_{2}}(x_{\B},p_{\B}).
\end{align}
Here, $\{\rho_{n_{1},n_{2};\ti{\bf  n}}\}$ are operators in the original system subspace, and
$\rho_{00;\ti{\bf  0}}$ is just the reduced system operator.
The scaling parameters, $\{s_{n_1,n_2}\}$, and the functions $\{\Psi_{n_{1},n_{2}}\}$ are detailed in Appendix \ref{FP}.
After some simple algebra, we obtain $\rho_{{\bf n};{\ti{\bf n}}} \equiv \rho_{n_{1},n_{2};\ti{\bf  n}}$ the EOM,
\begin{align}\label{hdeom}
	\dot \rho_{{\bf n};{\ti{\bf n}}} = & -\big(i{\cal L}_{\tP} +\gamma_{\,\ti{\bf  n}} \big ) \rho_{{\bf n};{\ti{\bf n}}} -i \sum_{k=1}^{N_K}\hat x_{\B}^{\times}\rho_{{\bf n};{\ti{\bf  n}}_{k}^+}
	\nl                               &
	-i\sum_{k=1}^{N_K} {\bm \ti n}_{k} \Big(
	\ti\eta_{k}\hat x_{\B}^{\greater}\rho_{{\bf n};{\ti{\bf  n}}_{k}^-}
	-\ti\eta^{\ast}_{\bar k} \hat x_{\B}^{\lesser}\rho_{{\bf n};{\ti{\bf  n}}_{k}^-}
	\Big).
\end{align}
The solvation mode actions, $x_{\B}^{\times}$ and $\hat x_{\B}^{\lgter}$ 
are considered in the Wigner representation; see \Eqs{app_xleftaction}--(\ref{app_prightaction}).
Moreover,
\begin{align} \label{L-core}
	{\cal L}_{\tP} \rho_{{\bf n};{\ti{\bf n}}} &
	=  {\cal L}_{\tS} \rho_{{\bf n};{\ti{\bf n}}} + \hat Q_{\tS}F(\hat x_{\B}^{\greater})\rho_{{\bf n};{\ti{\bf n}}}-F(\hat{x}_{\B}^{\lesser})\rho_{{\bf n};{\ti{\bf {n}}}}\hat Q_{\tS}
	\nl                                       & \quad
	+ \frac{\w_{\B}}{2}\big(\hat p_{\B}^{\greater 2}+\hat x_{\B}^{\greater 2}-\hat p_{\B}^{\lesser 2}-\hat x_{\B}^{\lesser 2}\big) \rho_{{\bf n};{\ti{\bf n}}}
	\nl                                       & \quad
	+{\ti \lambda} \big(\hat x_{\B}^{\greater 2}-\hat x_{\B}^{\lesser 2}\big) \rho_{{\bf n};{\ti{\bf n}}} .
\end{align}
The above solvation mode subspace Wigner representation also highlights the  role of the FP formulation.\cite{Wig32749,Ris89,Xu17395,Liu18245}

\subsection{Numerical demonstrations}\label{sec_fp_b}

\begin{figure}[t]
	\centering
	\includegraphics{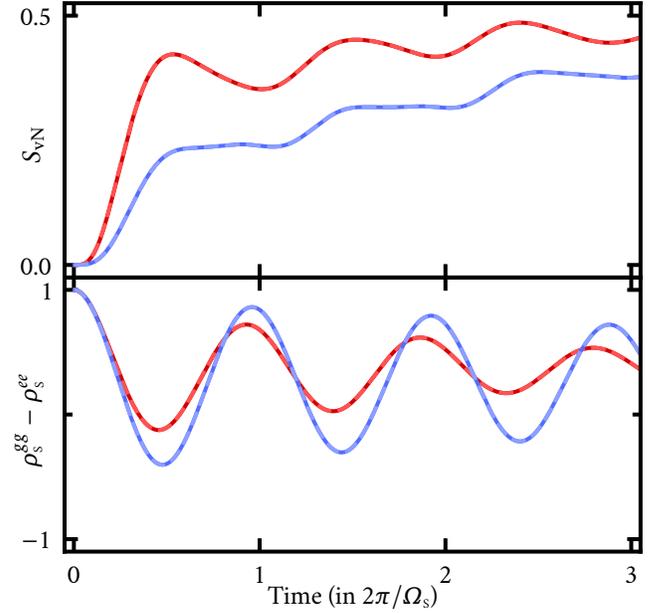}
	\caption{Comparison between extended DEOM (solid) and BSM-HQME (dashed) in terms of von Neumann entropy, $S_{\mathrm{vN}} = -\mathrm{tr}_{\tS} (\rho_{\tS} \ln \rho_{\tS})$, and the evolution of population, $\rho_{\mbox{\tiny S}}^{gg} - \rho_{\mbox{\tiny S}}^{ee}$.
		Here, we set $\w_{eg} = V = \w_{\tB}$. The Rabi frequency is $\Omega_{\tS} = \sqrt{5} \w_{\tB}$.
		The bath related parameters are selected as $(\theta_{\B}$, $\lambda / \w_{\tB})$ $=$ $(1.125$, $0.1)$ and $(1.333$, $0.25)$, plotted in blue and red curves, respectively.
		%
		%
		%
	}
	\label{fig1}
\end{figure}

\begin{figure}[htbp]
	\centering
	\includegraphics{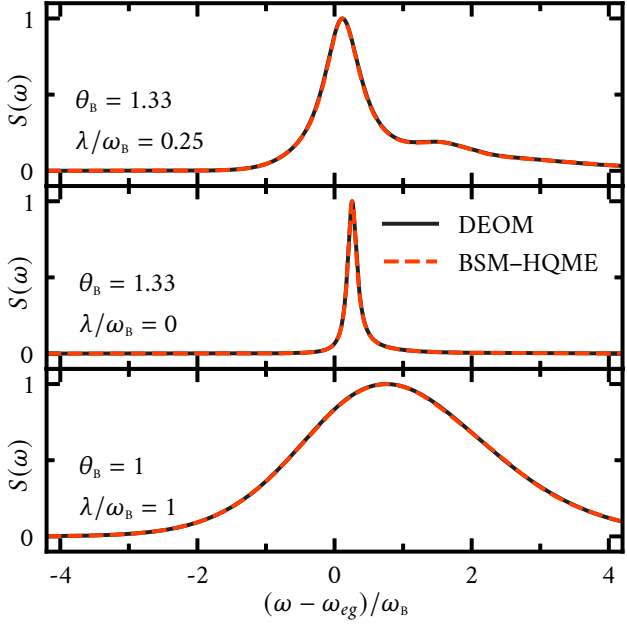}
	\caption{
		Absorption spectra via the extended DEOM (black solid) and the BSM-HQME (red dashed).
		We set $\w_{eg} = 50 \w_{\tB}$ and three pairs of the specified bath related parameters $(\theta_{\B}$, $\lambda / \w_{\tB})$. Other parameters are the same as those in \Fig{fig1}.
	}
	\label{fig2}
\end{figure}

\begin{figure*}[t!]
	\centering
	\includegraphics{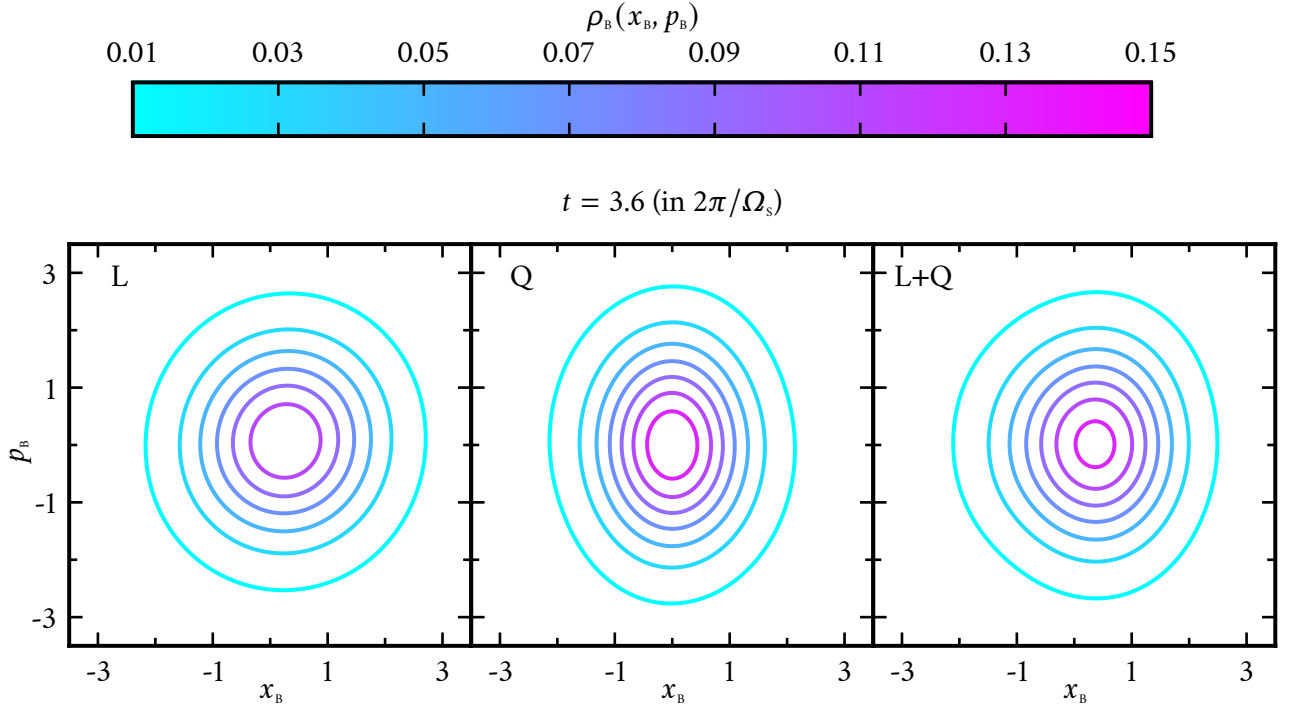}
	\caption{
		Phase space dynamics of solvation mode evaluated via BSM-HQME with $(\theta_{\B}, \lambda / \w_{\tB}) = $ $(1, 0.25)$, $(1.33, 0)$ and $(1.33, 0.25)$, representing the pure linear (L), the pure quadratic (Q) and the mixed (L$+$Q) coupling bath scenarios. Other parameters are the same as those in \Fig{fig1}. (Multimedia view)
	}
	\label{fig3}
\end{figure*}

For demonstration, we select a two--state model system as in Ref.\ \onlinecite{Xu18114103}. In this model, the total Hamiltonian can be described by
\be
H_{\T} = h_{g} |g \ra \la g | + (h_{e} + \w_{\eg}) |e \ra \la e |,
\ee
where
\begin{equation}
	h_{e} = \frac{1}{2} \w_{\tB} (\hat p_{\tB}^2 + \hat x_{\tB}^2) + \frac{1}{2} \sum_{k} \widetilde{\w}_{k} \Big[\tilde{p}_{k}^2 + \Big(\tilde{x}_k - \frac{\tilde{c}_{k}}{\widetilde{\w}_{k}} \hat{x} _{\tB}\Big)^2 \Big],
\end{equation}
and
\begin{equation}
	h_{g} = \frac{1}{2} \w'_{\tB} ({\hat p}_{\tB}^{\prime 2} + {\hat x}_{\tB}^{\prime 2}) + \frac{1}{2} \sum_{k} \widetilde{\w}'_{k} \Big[\tilde{p}_{k}^{\prime 2} + \Big(\tilde{x}'_k - \frac{\tilde{c}'_{k}}{\widetilde{\w}'_{k}} \hat{x}_{\tB}'\Big)^{2}\Big],
\end{equation}
follow the Caldeira--Leggett's interaction form. \cite{Cal83587}
Then in the $H_{g}$--based description, $H_{\T}$ can be reformulated as
\begin{align}
	H_{\T} = & \w_{\eg} |e \ra \la e| + h_{g} + (h_{e} - h_{g})|e \ra \la e|
	\nl =    & \w_{\eg} |e \ra \la e| + h_{g} + (\alpha_0 + \alpha_1 \hat x_{\tB} + \alpha_2 {\hat x}^2_{\tB})|e \ra \la e|,
\end{align}
with
\begin{equation}
	\alpha_0=\lambda\theta^2_{\B},
	\quad
	\alpha_1=-(2\lambda\omega_{\B})^{\frac{1}{2}}\theta^2_{\B},
	\quad
	\alpha_2=\frac{\omega_{\B}}{2}(\theta^2_{\B}-1).
\end{equation}
Here, $\lambda$ represents the linear--displacement induced reorganization and $\theta_{\tB} = \w'_{\tB} / \w_{\tB}$. Both the linear and quadratic coupling strengths are determined by these two parameters.
We adopt the Drude model for the secondary bath,
\be\label{hatzetaB}
\zeta_{\B}(\w)=\frac{2 i \ti \lambda \w_{\B}}{\w + i \ti \gamma},
\ee
with $\ti \lambda$ being the reorganization energy defined in \Eq{ti_lambda}. The resulted $\ti J(\w)$ reads [cf.\ \Eq{spec2nd}]
\be
\ti J(\w)=\frac{2 \ti \lambda \ti\gamma\w}{\w^2+\ti\gamma^2}.
\ee
We set $\ti \lambda = 5\w_{\tB}$, $\ti\gamma = 15\w_{\tB}$ and $\beta \w_{\tB} = 1$, with $\w_{\tB}$ being the unit in the following demonstrations.

Figure \ref{fig1} shows the extended DEOM dynamics, in comparison with the BSM-HQME results, on the von Neumann entropy (upper panel) and the population (lower panel).
We assume the system Hamiltonian as $H_{\tS} = \w_{\eg} | e \ra \la e | + V (| e \ra \la g | + | g \ra \la e |)$ and the initial state being the equilibrium ground state in the absence of nonadiabatic coupling ($V = 0$).
As seen from the figure, the extended DEOM results agree perfectly with those of the BSM-HQME.
We also carry out the SFD--DEOM \cite{Che21174111} calculation with the parameters $(\theta_{\B}$, $\lambda / \w_{\tB})$ $=$ $(1.125$, $0.1)$. The converged results (not shown in the figure) also match those of both BSM-HQME and extended DEOM.
The SFD--DEOM encounters error accumulation in the longtime simulation, requiring a vast of trajectories to converge when $\alpha_{2}$ is relatively large.
%
%
In the BSM-HQME simulations, we exploit 
the  on--the--fly numerical filter technique, \cite{Shi09084105} with the accuracy of  $10^{-8}$ that effectively  corresponds to the converged level of  truncation at  $n_1 + n_2=8 $ and $\sum_k { \ti n}_k = 30$.

Figure \ref{fig2} reports the evaluated absorption spectra using the extended DEOM and the BSM-HQME, with three pairs of the specified bath parameters $(\theta_{\B}$, $\lambda / \w_{\tB})$.
The absorption spectrum is defined as
\begin{equation}
	\label{absorp}
	S(\w) = \mathrm{Re} \int _{0}^{\infty} \d t e^{i \w t} \la \hat \mu_{\tS}(t) \hat \mu_{\tS}(0) \ra
\end{equation}
with $\hat \mu_{\tS} = |e \ra \la g| + |g \ra \la e|$.
As shown in the figure, the extended DEOM and BSM-HQME agree with each other perfectly.

Figure \ref{fig3} (Multimedia view) depicts the time evolution of the solvation mode phase--space distribution [cf.\ \Eq{expansion} with \Eqs{explicit_basis} and (\ref{psi_single})],
\begin{equation}
	\rho_{\B}(x_{\B},p_{\B}; t) \equiv \mathrm{tr}_{\tS} [W_{\bf \ti 0}(x_{\B},p_{\B}; t)].
\end{equation}
We choose $(\theta_{\B}, \lambda / \w_{\tB}) = $ $(1, 0.25)$, $(1.33, 0)$ and $(1.33, 0.25)$ to represent the pure linear (L), the pure quadratic (Q) and the mixed (L$+$Q) coupling bath scenarios, respectively.
The video shows the dynamic interplay between the nonadiabatic coupling $V$, the linear bath coupling that contributes to the center displacement, and the quadratic bath coupling that causes the curvature change.
Evidently, compared to the DEOM, the BSM-HQME is the choice to ``visualize'' the solvation mode dynamics.

\section{Thermodynamic mixing}\label{sec_therm}
In this section, we extend the present DEOM formalism to thermodynamics problems. Involved would be the imaginary--time DEOM ($i$-DEOM) and also nonequilibrium $\lambda$-DEOM (neq-$\lambda$-DEOM), to be detailed in \Sec{i-deom} and \Sec{lambda-deom}, respectively. The key quantities are the system--bath mixing free--energy, nonequilibrium work and its distribution function. The numerical results in \Sec{numerical}  reproduce the Jarzynski equality \cite{Jar11329} and the Crooks relation. \cite{Cro992721}

\subsection{Extended $i$-DEOM formalism}
\label{i-deom}
The $i$-DEOM aims at the hybridization partition function,
\be \label{uu2}
Z_{\rm hyb}\equiv Z_{\T}/Z_{0}
\equiv {\rm Tr} \varrho_{\T} (\beta),
\ee
with $Z_{\T}\equiv{\rm Tr}\,e^{-\beta H_{\T}}$ and $Z_{0}\equiv{\rm Tr}\,e^{-\beta H_{0}}$, where $H_{0} = H_{\tS} + h_{\B}$ [cf.\ \Eq{HT0}].
The hybridization free--energy before and after isotherm system--bath mixing, $A_{\rm hyb}(T)\equiv A_{\T}(T)- A_0(T)$, is given by
\begin{equation} \label{free_enegy}
	A_{\rm hyb}(T)=-\beta^{-1}\ln Z_{\rm hyb}(T).
\end{equation}
This is to be evaluated by using \Eq{uu2} via
\be\label{varrhoT_tau}
\varrho_{\T}(\tau) = e^{-\tau H_{\T}}e^{-(\beta-\tau)H_{0}}\big / Z_0,
\ee
that satisfies
\begin{equation}
	\label{i-deom-equation}
	\frac{\d \varrho_{\T}(\tau)} {\d\tau} = -(H_{\tS}^{\times} + h_{\B}^{\times} + H_{\SB}^{>}) \varrho_{\T}(\tau),
\end{equation}
where $H_{\SB} \equiv \hat Q_{\tS}(\alpha_0+\alpha_1\hat x_{\B}+ \alpha_2\hat x_{\B}^{2})$.
The $i$-DEOM--space mapping goes by \cite{Gon20214115, Wan22170901}
\be \label{i-ddos}
\varrho_{\T}(\tau)\rightarrow
{\bm\varrho}(\tau) \equiv \{\varrho^{(n)}_{\bf n}(\tau)\}.
\ee
We obtain \cite{Gon20214115, Wan22170901}
\begin{equation}
	\label{iDEOMa}
	\frac{\d \varrho^{(n)}_{\bf n} (\tau)} {\d\tau}  = - \varrho^{(n)}_{\bf n} (\tau; H_{\tS}^{\times}) - \varrho^{(n)}_{\bf n} (\tau; h_{\B}^{\times}) - \varrho^{(n)}_{\bf n} (\tau; H_{\SB}^{>})
\end{equation}
with
\begin{align}
	\varrho^{(n)}_{\bf n} (\tau; H_{\SB}^{>}) & = \alpha_0 \hat Q_{\tS} \varrho^{(n)}_{\bf n} (\tau) + \alpha_1 \hat Q_{\tS} \varrho^{(n)}_{\bf n} (\tau; \hat x_{\B}^{>}) \nl
	                                          & \quad + \alpha_2 \hat Q_{\tS} \varrho^{(n)}_{\bf n} (\tau; \hat x_{\B}^{>2}).
\end{align}

In parallel to the dissipaton algebra introduced in \Sec{sec_eDEOM}, especially \Eq{Wick_xB2} that deal with the quadratic coupling, we obtain the final $i$-DEOM formalism,
\begin{align} \label{i_DEOM}
	\dot\varrho^{(n)}_{\bf n} \!
	= & \Big(-{\cal L}_{\tS} + i \sum_{k}n_k\gamma_k\Big) \varrho^{(n)}_{\bf n} - \left(\alpha_0+\alpha_2\la\hat x^2_{\B}\ra_{\B}\right) {\bar{\cal A}} \varrho^{(n)}_{\bf n} \nl
	  & - \alpha_1 \sum_{k} \Big({\bar{\cal A}} \varrho^{(n+1)}_{{\bf n}^{+}_{k}} + n_k{\bar{\cal C}}_k \varrho^{(n-1)}_{{\bf n}^{-}_{k}} \Big) - 2 \alpha_2 \sum_{kk'} n_k {\bar{\cal C}}_k \varrho^{(n)}_{{\bf n}^{-+}_{kk'}}\nl
	  & - \alpha_2 \sum_{kk'} \Big[{\bar{\cal A}} \varrho^{(n+2)}_{{\bf n}^{++}_{kk'}} + n_k (n_{k'}-\delta_{kk'}) {\bar{\cal B}}_{kk'} \varrho^{(n-2)}_{{\bf n}^{--}_{kk'}} \Big],
\end{align}
with
\bsube\label{barcalABC}
\begin{align}
	 & {\bar{\cal A}}\hat O \equiv \hat Q_{\tS} \hat O,\quad {\bar{\cal C}}_k\hat O \equiv \eta_k \hat Q_{\tS}\hat O.
	\\
	 & {\bar{\cal B}}_{kk'} \hat O \equiv \eta_k \eta_{k'} \hat Q_{\tS} \hat O.
\end{align}
\esube
Evidently, when $\alpha_{2} = 0$, \Eq{i_DEOM} reduces to the conventional $i$-DEOM formalism.\cite{Gon20214115, Wan22170901}
The solutions of $Z_{\rm hyb} = {\rm Tr} \varrho_{\bf 0}^{(0)} (\beta)$ can be obtained by propagation of \Eq{i_DEOM} from the initial values $\varrho_{\bf 0}^{(0)} (0) = e^{-\beta H_{0}} / Z_0$ and $\varrho_{\bf n}^{(n > 0)} (0) = 0$.

\subsection{Extended neq-$\lambda$-DEOM formalism}
\label{lambda-deom}
Turn to the neq-$\lambda$-DEOM formalism, aiming at the system--bath mixing nonequilibrium work and its distribution function.\cite{Gon22054109, Wan22170901} The involved $\lambda(t)$--augmented total composite Hamiltonian reads
\be \label{Hlam}
H_{\T}(t) = H_{\tS} + h_{\B}  + \lambda(t) H_{\tS\B}.
\ee
A time--dependent mixing function $\lambda(t)$ is used subject to $\lambda(t=0)=0$ and $\lambda(t=t_f) = 1$.
This represents a nonequilibrium scenario. In related studies, the work distribution $p(w)$ is the key quantity.
There exists the Jarzynski equality \cite{Jar11329}
\be \label{Jar}
\La e^{-\beta w}\Ra\equiv\int_{-\infty}^{\infty}\!\!{\rm d} w \,e^{-\beta w} p(w)=e^{-\beta A_{\rm hyb}}
\ee
and the Crooks relation \cite{Cro992721}
\be\label{CrooksEq}
e^{-\beta w} p(w) = e^{-\beta A_{\rm hyb}} \bar{p}(-w).
\ee
The latter is about a pair of conjugate processes, with the forward and backward processes being controlled by $\lambda(t)$ and $\bar\lambda(t)\equiv\lambda(t_f-t)$, respectively.\cite{Tal07F569}
The forward work distribution $p(w)$ and the backward $\bar p(-w)$ cross at $w=A_{\rm hyb}$ where $A_{\rm hyb}$ can be obtained via the $i$-DEOM calculation with \Eqs{uu2} and (\ref{free_enegy}).

The neq-$\lambda$-DEOM enables the accurate evaluation of $p(w)$.
To proceed, we start with  $H_{0}|n\ra=H_{\T}(\lambda=0)|n\ra=\varepsilon_n|n\ra$ and $H_{\T}(\lambda=1)|N\ra=E_N|N\ra$ before and after mixing.
The distribution of mixing work is given by\cite{Tal07050102}
\be\label{forwardfW}
p(w) = \sum_{N,n}\delta(w - E_N + \varepsilon_n)
P_{N,n}(t_f, 0) e^{- \beta \varepsilon_{n}} / {Z_0}.
\ee
In \Eq{forwardfW}, $P_{N,n}(t,0) = \big \vert \la N \vert \hat U_{\T} (t) \vert n \ra \big \vert^2$ is the transition probability with the propagator $\hat U_{\T}(t)$ being governed by the Hamiltonian $H_{\T}(t) = H_{0} + \lambda(t) H_{\SB}$.

Define then
\be\label{Phitau1}
\hat \Phi_{\T}(t;\tau) =
\hat U_{\T}(t)\hat V_{+}(t;\tau) \rho^{\rm eq}_{0}(T)\hat V_{-}(t;\tau)\hat U_{\T}^{\dg}(t),
\ee
where
\begin{align}\label{Vpm}
	\hat V_{\pm}(t;\tau) 
	=\exp_{\pm}\left[\frac{i\tau}{2}\!\int_{0}^{t}{\rm d}t'\, \dot{\lambda}(t')H_{\SB}\right].
\end{align}
It can be shown that\cite{Tal07050102,Sak21033001}
\be\label{varPhitau1}
p(w)=\frac{1}{2\pi}\!\int_{-\infty}^{\infty}\!\!{\rm d} \tau \,e^{-i w\tau } {\rm Tr}\hat \Phi_{\T}(t_f;\tau).
\ee
This concludes that $\hat \Phi_{\T}(t;\tau)$, defined in \Eq{Phitau1}, is the work generating operator, satisfying
\begin{align}\label{eom}
	\frac{\partial \hat \Phi_{\T}} {\partial t} & = -i[H_{\tS}^{\times} + h_{\B}^{\times} + \lambda_{-}(t) H_{\SB}^{\greater} - \lambda_{+}(t) H_{\SB}^{\lesser}]\hat \Phi_{\T},
\end{align}
with
\begin{equation}
	\label{def_lambda_tau}
	\lambda_{\pm}(t)\equiv \lambda(t)\pm(\tau/2)\dot\lambda(t).
\end{equation}
Initially, $\hat \Phi_{\T}(0;\tau)=\rho^{\rm eq}_{0}(T)=e^{-\beta H_{0}}/Z_0$, as inferred from  \Eq{Phitau1}.

Similar to \Eq{i-ddos}, we obtain the dissipatons--augmented work generating operators (D-WGOs) mapping,
\begin{equation}
	\hat \Phi_{\T}(t;\tau)\rightarrow \hat{{\bm \Phi}}(t;\tau) \equiv \{\hat \Phi^{(n)}_{\bf n}(t;\tau)\}.
\end{equation}
Following the similar procedure from \Eq{SchEq} to \Eq{gen_DEOM} applied to \Eq{eom}, we obtain the D-WGO correspondence.\cite{Gon22054109, Wan22170901}

\begin{align} \label{lambda_DEOM}
	\dot \Phi^{(n)}_{\bf n} \!
	= & - \Big(i {\cal L}_{\tS} + \sum_{k}n_k\gamma_k\Big) \Phi^{(n)}_{\bf n} - i \Big(\alpha_0+\alpha_2\la\hat x^2_{\B}\ra_{\B}\Big) {\ti{\cal A}}(t) \Phi^{(n)}_{\bf n} \nl
	  & - i \alpha_1 \sum_{k} \Big[{\ti{\cal A}}(t) \Phi^{(n+1)}_{{\bf n}^{+}_{k}} + n_k{\ti{\cal C}}_k (t) \Phi^{(n-1)}_{{\bf n}^{-}_{k}}\Big] \nl
	  & - i \alpha_2 \sum_{kk'} \Big[{\ti{\cal A}} (t) \Phi^{(n+2)}_{{\bf n}^{++}_{kk'}} + n_k(n_{k'}-\delta_{kk'}){\ti{\cal B}}_{kk'} (t) \Phi^{(n-2)}_{{\bf n}^{--}_{kk'}}\Big] \nl
	  & - 2 i \alpha_2 \sum_{kk'} n_k {\ti{\cal C}}_k (t) \Phi^{(n)}_{{\bf n}^{-+}_{kk'}},
\end{align}
where
\bsube
\label{ticalABC}
\begin{align}
	 & {\ti{\cal A}} (t) \hat O \equiv \lambda_{-}(t) \hat Q_{\tS} \hat O - \lambda_{+}(t) \hat O \hat Q_{\tS},
	\\
	 & {\ti{\cal B}}_{kk'} (t) \hat O \equiv \lambda_{-}(t) \eta_{k} \eta_{k'} \hat Q_{\tS} \hat O
	- \lambda_{+}(t) \eta^{\ast}_{\bar k} \eta^{\ast}_{\bar k'}\hat O \hat Q_{\tS},
	\\
	 & {\ti{\cal C}}_k (t) \hat O \equiv \lambda_{-}(t) \eta_k \hat Q_{\tS}\hat O
	- \lambda_{+}(t) \eta^{\ast}_{\bar k}\hat O\hat Q_{\tS}.
\end{align}
\esube
In relation to $\hat \Phi_{\T}(0;\tau)=\rho^{\rm eq}_{0}(T)=e^{-\beta H_{0}}/Z_0$, the initial values to \Eq{lambda_DEOM} are
\be\label{iDDOs0}
\hat \Phi^{(0)}_{\bf 0}(0;\tau)=e^{-\beta H_{\tS}}/Z_{\tS}
\ \ \text{and} \ \
\hat \Phi_{\bf n}^{(n>0)}(0;\tau)=0.
\ee
Finally, by using \Eq{lambda_DEOM} with \Eqs{ticalABC} and (\ref{def_lambda_tau}), we evaluate
\be\label{pww}
p(w)=\frac{1}{2\pi}\int_{-\infty}^{\infty}\!\!{\rm d} \tau \,e^{-i w\tau } {\rm tr}_{\tS}[\hat \Phi^{(0)}_{\bf 0}(t_f;\tau)].
\ee

\subsection{Numerical demonstrations}
\label{numerical}
\begin{figure}[t]
	\centering
	\includegraphics{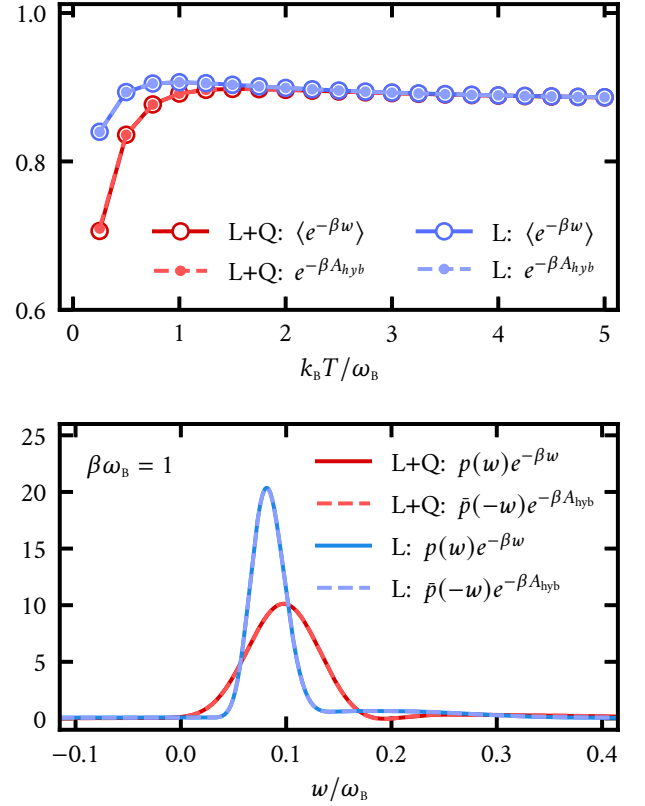}
	\caption{
		The results of DEOM with GWT-2, in terms of the Jarzynski equality (upper panel) and the Crooks relation (lower panel).
		Parameters are the same as those of \Fig{fig3}.
		%
		%
	}
	\label{fig4}
\end{figure}

In our simulations, we set the forward and backward time-dependent mixing function to be
\begin{equation}
	\lambda (t) = \frac{1 - e^{-a t}}{1 - e^{-a t_{f}}},
\end{equation}
and
\begin{equation}
	\bar \lambda (t) = \lambda (t_f - t) = \frac{e^{a t_{f}} - e^{a t}}{e^{a t_{f}} - 1},
\end{equation}
respectively.
Figure \ref{fig4} reports the DEOM results in terms of the Jarzynski equality (\ref{Jar}) and the Crooks relation (\ref{CrooksEq}).
The $e^{-\beta A_{\rm hyb}}$ and the $w$ related functions are evaluated by the methods described in \Sec{i-deom} and \Sec{lambda-deom}, respectively.
Numerical simulations are carried out with the same setup as in \Fig{fig3}.
The upper panel shows that the Jarzynski equality is recovered. At the high--temperature regime, the classical equipartition theorem holds as seen from the coincidence between the results of L and L$+$Q scenarios.
The lower panel reproduces the Crooks relation. The resulting work distribution function can be used in analyzing the cumulants of various orders, including the center, the variance, the skewness and so on. \cite{Esp091665}
%
%
It is worth reemphasizing that \Fig{fig4} in turn confirms the rigorousness of the extended DEOM theories, covering the real--time, the imaginary--time and the nonequilibrium $\lambda(t)$ dynamics.


\section{Summary}\label{sec_sum}
To conclude, we present a comprehensive account of extended DEOM with quadratic environments, which could in principle be generalized to arbitrary nonlinear bath coupling scenarios.
The developments include also an equivalent core-system phase--space hierarchy construction, BSM-HQME, as verified both theoretically and numerically.
The extended DEOM is numerically more efficient, whereas the core--system BSM-HQME is favorable for ``visualizing'' the correlated solvation dynamics.

While the present theories are elaborated with the single dissipative mode case, the extensions to multiple dissipative modes, including modes mixing (i.e.\ Duschinsky rotation), are straightforward.
Moreover, the existing system--bath entanglement theorem with linear environment coupling \cite{Du20034102, Gon20214115, Wan22044102, Wan22170901} can be further investigated with the inclusion of quadratic environment coupling.
The aforementioned theoretical developments would comprise required toolkits for the construction of practical dissipaton--correlated coarse--graining molecular dynamics and thermodynamics methods.

It is also worth noting that the GWTs are validated within the canonical  Feynman--Vernon influence functional formalism with the linear environmental couplings. \cite{Wan20} 
However,  it is rather  cumbersome to prove the GWTs and derive the EOM in nonlinear  coupling scenarios using the Feynman--Vernon influence functional approach.
There is no simple analytical expression of nonlinear influence functionals.
On the other hand, the dissipaton algebra  enables the GWT-2 in the most direct manner. The resulting \Eq{gen_DEOM}, the extended DEOM formalism,\cite{Xu17395, Xu18114103} is numerically validated by comparing with other two very different approaches, BSM-HQME and SFD-DEOM, in this work.
This indicates that the dissipaton theory would be an important building block towards the future development of
 open quantum systems.

\begin{acknowledgements}
	Support from the Ministry of Science and Technology of China (Grant No.\ 2021YFA1200103) and the National Natural Science Foundation of China (Grant Nos.\ 22103073 and 22173088) is gratefully acknowledged.
\end{acknowledgements}

\appendix*
\section{Onset of  algebra for  Brownian solvation mode}\label{FP}%
\subsection{The Fokker--Planck operator and basis set construction}\label{sec:fp}
The following part is the procedure to generate the FP basis set for describing the solvation mode inside the core system.
Let us start from \Eq{FDT} with \Eq{JBO}. In the Markovian limit, $\zeta_{\B}(\w)=\zeta_{\B}$, by further adopting the {high--temperature} approximation, \cite{Din17024104} we have
\be\label{double_exp}
\la\hat x_{\B}(t)\hat x_{\B}(0)\ra_{\B}\approx \eta_{+} e^{-\gamma_+ t}+\eta_-e^{-\gamma_- t},
\ee
with
\be
\gamma_{\pm}=\frac{1}{2}[\zeta_{\B}\pm(\zeta_{\B}^2-4\w_{\B}^2)^{\frac{1}{2}}],
\ee
and
\be \label{etaHT}
\eta_{\pm}=\mp\frac{\w_{\B}}{\gamma_{+}-\gamma_{-}}\Big(\frac{1}{\beta\gamma_{\pm}}-\frac{i}{2}\Big).
\ee
It should be noticed that the Markovian and high--temperature limits are needed only for the construction of FP operator and FP algebra, which are \emph{not} necessary for the BSM-HQME, which is exact at any temperature.

Then one may construct the FP operator ${\cal L}_{\tFP}$ that would generate the FP basis set,\cite{Ris89,Xu09074107,Din17024104}
\begin{align}\label{FP_op}
	{\cal L}_{\tFP}(\cdot)\equiv & \ i\frac{\w_{\B}}{2}[\hat p_{\B}^2+\hat x_{\B}^2,\cdot]+\frac{\zeta_{\B}}{\beta\w_{\B}}[\hat x_{\B},[\hat x_{\B},\cdot]]
	\nl                          &
	+i\frac{\zeta_{\B}}{2}[\hat x_{\B},\{\hat p_{\B},\cdot\}].
\end{align}
In Wigner representation, the FP operator, \Eq{FP_op}, can be converted to a solvation phase--space operator as
\be\label{fpo}
L_{\tFP}\!\equiv\!\w_{\B}\Big(\frac{\partial}{\partial x_{\B}}p_{\B}\!-\frac{\partial}{\partial p_{\B}}x_{\B}
\!\Big)\!-\frac{\zeta_{\B}}{\beta\w_{\B}}\frac{\partial^2}{\partial p_{\B}^2}-\zeta_{\B}\frac{\partial}{\partial p_{\B}}p_{\B},
\ee
where $x_{\B}$ and $p_{\B}$ are c--number valued solvation coordinate and momentum, respectively.
Under the similarity transformation
\be \label{tifpo}
\ti L_{\tFP} \equiv e^{S}L_{\tFP}e^{-S},
\ee
with $S\equiv\frac{\beta\w_{\B}}{4}(x_{\B}^2+p_{\B}^2)$, we can recast $\ti L_{\tFP}$ as
\begin{align} \label{tifpo2}
	\ti L_{\tFP} & =\w_{\B}(a_2^{\dg} a_1-a_1^{\dg}a_2)+\zeta_{\B}a_2^{\dg}a_2,
	\nl          &
	=\gamma_{+}c_1^{+}c_1^{-} +\gamma_{-}c_2^{+}c_2^{-}.
\end{align}
The first line involves solvation phase--space operators,
\bsube
\be
\begin{split}\label{aadg}
	&a_1=\frac{\sqrt{\beta\w_{\B}}}{2}x_{\B}+\frac{1}{\sqrt{\beta\w_{\B}}}\frac{\partial}{\partial x_{\B}}, \\
	&a_2=\frac{\sqrt{\beta\w_{\B}}}{2}p_{\B}+\frac{1}{\sqrt{\beta\w_{\B}}}\frac{\partial}{\partial p_{\B}},
\end{split}
\ee
and the corresponding
\be
\begin{split}\label{aadg}
	& a_1^{\dg}=\frac{\sqrt{\beta\w_{\B}}}{2}x_{\B}-\frac{1}{\sqrt{\beta\w_{\B}}}\frac{\partial}{\partial x_{\B}}, \\
	& a^{\dg}_2=\frac{\sqrt{\beta\w_{\B}}}{2}p_{\B}-\frac{1}{\sqrt{\beta\w_{\B}}}\frac{\partial}{\partial p_{\B}}.
\end{split}
\ee
\esube
In \Eq{tifpo2}, the $\w_{\B}$--term describes the coherent dynamics and the $\zeta_{\B}$--term is responsible for dissipation.
The second line describes the quasi-particle diagonalization, with
\be
\begin{split} \label{ccdg}
	&c_1^{-}\equiv r_2a_1+r_1a_2, \ \ \ \ \ \ c_2^{-}\equiv r_1a_1+r_2a_2,\\
	&c_1^{+}\equiv -r_2a_1^{\dg}+r_1a_2^{\dg}, \ \ \ \ c_2^{+}\equiv r_1a_1^{\dg}-r_2a_2^{\dg},
\end{split}
\ee
and
\be\label{r12}
r_1 \equiv \sqrt{\gamma_{+}/(\gamma_{+}-\gamma_{-})},\quad
r_2 \equiv \sqrt{\gamma_{-}/(\gamma_{+}-\gamma_{-})}.
\ee
Note that $c_{1,2}^{+}\neq (c_{1,2}^{-})^{\dg}$. 
It is easy to verify that they satisfy the bosonic commutation relations as
\be
\begin{split}\label{bosecommutation}
	&[c_j^{-},c_{j'}^{+}]=[a_j, a_{j'}^{\dg}]=\delta_{jj'},\\
	&[c_j^{\pm},c_{j'}^{\pm}]=[a_j, a_{j'}]=[a_j^{\dg}, a_{j'}^{\dg}]=0.
\end{split}
\ee

The eigenfunctions of $\ti L_{\tFP}$ can now be obtained as
\be\label{psi}
\Psi_{n_{1},n_{2}}(x_{\B},\!p_{\B})\!=\!\!\frac{1}{\sqrt{n_1!n_2!}}(c_1^{+})^{n_1}\!(c_2^{+})^{n_2}\Psi_{00}(x_{\B},p_{\B}),
\ee
with
\be\label{psi00}
\Psi_{00}(x_{\B},p_{\B})\equiv \left(\frac{\beta\w_{\B}}{2\pi}\right)^{\frac{1}{2}}e^{-\frac{\beta\w_{\B}}{4}(x_{\B}^2+p_{\B}^2)}.
\ee
It can be verified that
\be\label{eign}
\ti L_{\rm FP}\Psi_{n_{1},n_{2}}=(n_1\gamma_{+}+n_2\gamma_{-})\Psi_{n_{1},n_{2}}.
\ee
Furthermore, by denoting $\bar c_{j}^{\pm}\equiv (c_j^{\mp})^{\dg}$, we can also construct the eigenfunctions $\{\bar \Psi_{n_{1},n_{2}}\}$ of  $ \ti L_{\tFP}^{\dg}$ in a similar way, reading
\be \label{psibar}
\bar \Psi_{n_{1},n_{2}}(x_{\B},\!p_{\B})\!=\!\!\frac{1}{\sqrt{n_1!n_2!}}(\bar c_1^{+})^{n_1}\!(\bar c_2^{+})^{n_2}\Psi_{00}(x_{\B},p_{\B}),
\ee
satisfying
\be\label{eignbar}
\ti L_{\tFP}^{\dg}\bar\Psi_{n_{1},n_{2}}=(n_1\gamma_{+}^{\ast}+n_2\gamma_{-}^{\ast})\bar\Psi_{n_{1},n_{2}}.
\ee
The functions $\{\bar \Psi_{n_{1},n_{2}}\}$ are orthonormal with respect to $\{\Psi_{n_{1},n_{2}}\}$ as
\be \label{orth}
\int\!\!\!\!\int \!\!{\rm d}x_{\B}{\rm d}p_{\B}\bar \Psi^{\ast}_{\!n_{1}n_{2}}\!(x_{\B},p_{\B})\Psi_{\!n'_{1},n'_{2}}\!(x_{\B},p_{\B})\!\!=\!\delta_{n_1\!n_1'\!}\delta_{n_2n_2'\!}.
\ee

Any function $W(x_{\B},p_{\B})$, including that of \Eq{expansion}, can be expanded under this set of eigenfunctions  $\{\Psi_{n_{1},n_{2}}\}$ as
%
\begin{align}\label{doublerho}
	W(x_{\B},p_{\B}) = & \Big(\frac{\beta\w_{\B}}{2\pi}\Big)^{\frac{1}{2}} \sum_{n_{1},n_{2}} \frac{s_{n_{1},n_{2}}}{\sqrt{n_1!n_2!}} \rho_{n_{1},n_{2}}
	\nl                &
	\times  e^{-\frac{\beta\w_{\B}}{4}(x_{\B}^2+p_{\B}^2)} \Psi_{n_{1},n_{2}}(x_{\B},p_{\B}).
\end{align}
Here, the parameters
\be
s_{n_{1},n_{2}}\equiv(-1)^{n_1}(\beta\w_{\B})^{\frac{n_{1} + n_{2}}{2}}r_2^{-n_{1}}r_1^{-n_{2}}
\ee
are chosen for later convenience and $\{\rho_{n_{1},n_{2}}\}$ are the coefficients. It should be noted that  $\{\rho_{n_{1},n_{2}}\}$  can be not necessarily c-numbers, but operators including other degrees of freedom.
Inversely, using the orthonormal relation, \Eq{orth}, we have
\begin{align}
	\rho_{n_{1},n_{2}}= & \frac{\sqrt{n_1!n_2!}}{s_{n_{1},n_{2}}}\left(\frac{\beta\w_{\B}}{2\pi}\right)^{\!\!-\frac{1}{2}} \int\!\!\!\int {\rm d}x_{\B}{\rm d}p_{\B} \bar \Psi^{\ast}_{n_{1}n_{2}}(x_{\B},p_{\B})
	\nl                 &
	\times e^{\frac{\beta\w_{\B}}{4}(x_{\B}^2+p_{\B}^2)}W(x_{\B},p_{\B}).
\end{align}
Moreover, the $\Psi_{n_{1}n_{2}}(x_{\B},p_{\B})$ can be explicitly expressed via \Eq{psi} with \Eqs{aadg} and (\ref{ccdg}), as
\begin{align}\label{explicit_basis}
	\Psi_{n_{1}n_{2}}(x_{\B},p_{\B})= & \sum_{j_1,j_2}{n_1 \choose j_1}{n_2 \choose j_2}r_1^{n_2-j_2+j_1}(-r_2)^{n_1-j_1+j_2}
	\nl                               &
	\times \sqrt{\frac{(n_1+n_2-j_1-j_2)!(j_1+j_2)!}{n_1!n_2!}}
	\nl                               &
	\times \psi_{n_1+n_2-j_1-j_2}(x_{\B})\psi_{j_1+j_2}(p_{\B}).
\end{align}
Here, the harmonic eigenfunctions, $\psi_{n}(z)$, read
\begin{align} \label{psi_single}
	\psi_{n}(z) \equiv \left(\frac{\beta\omega_{\B}}{2\pi}\right)^{\frac{1}{4}}\frac{e^{-\frac{\beta\omega_{\B}}{4}z^2}}{\sqrt{2^n n!}}H_n\left(\sqrt{\frac{\beta\omega_{\B}}{2}}z\right)
\end{align}
where $H_n(z)$ is the $n$th-order Hermitian polynomials.
Thus we finish the procedure to generate the FP basis set, $\{\Psi_{n_{1}n_{2}}\}$ [cf.\,\Eq{explicit_basis}].

\subsection{Actions of coordinate and momentum operators}\label{sec:xp}
In this subsection, we derive the actions of coordinate and momentum operators on $W(x_{\B},p_{\B})$ expressed in \Eq{doublerho}.
Firstly we map the left/right actions of coordinate and momentum operators $\hat x^{\gler}_{\B}$ and $\hat p^{\gler}_{\B}$ into the Wigner representation as\cite{Scu97}
\begin{align}
	\hat x_{\B}^{\gler}\rightarrow x_{\B}\pm\frac{i}{2}\frac{\partial}{\partial p_{\B}}
	\ \ \text{and} \ \
	\hat p_{\B}^{\gler}\rightarrow p_{\B}\mp\frac{i}{2}\frac{\partial}{\partial x_{\B}}.
\end{align}
Then by using \Eq{doublerho}, we obtain the solvation mode actions on $\rho_{n_{1},n_{2}}$ as
\begin{subequations} \label{app_xpaction}
	\begin{align}		\label{app_xleftaction}
		\hat x_{\B}^{\greater} \rho_{n_{1},n_{2}}                                  & = \rho_{n_{1}+1,n_{2}} + \rho_{n_{1},n_{2}+1} + n_1 \eta_{+} \rho_{n_{1}-1,n_{2}}
		\nl                                                                        & \ \ \ \
		+ n_2 \eta_{-}\rho_{n_{1},n_{2}-1},
		\\
		\label{app_xrightaction}\hat x_{\B}^{\lesser} \rho_{n_{1},n_{2}}           & = \rho_{n_{1}+1,n_{2}} + \rho_{n_{1},n_{2}+1} + n_1 \bar\eta_{+}^{\ast}\rho_{n_{1}-1,n_{2}}
		\nl                                                                        & \ \ \ \
		+ n_2 \bar \eta_{-}^{\ast} \rho_{n_{1},n_{2}-1},
		\\
		\label{app_pleftaction}
		\w_{\B} {\hat p}_{\B}^{\greater} \rho_{n_{1},n_{2}}                        & = -\gamma_{+}\rho_{n_{1}+1,n_{2}}\!-\!\gamma_{-}\rho_{n_{1},n_{2}+1} - n_1 \bar \eta_{-}^{\ast}\gamma_{-} \rho_{n_{1}-1,n_{2}}
		\nl                                                                        & \ \ \ \
		-n_2 \bar \eta_{+}^{\ast} \gamma_{+} \rho_{n_{1},n_{2}-1},
		\\
		\label{app_prightaction} \w_{\B}{\hat p}_{\B}^{\lesser} \rho_{n_{1},n_{2}} & = -\gamma_{+}\rho_{n_{1}+1,n_{2}} - \gamma_{-}\rho_{n_{1},n_{2}+1} - n_1 \eta_{-} \gamma_{-} \rho_{n_{1}-1,n_{2}}
		\nl                                                                        & \ \ \ \
		- n_2 \eta_{+} \gamma_{+} \rho_{n_{1},n_{2}-1}.
	\end{align}
\end{subequations}
Here, we denote
\be \label{etaHTbar}
\bar\eta_{\pm}^{\ast}\equiv\mp\frac{\w_{\B}}{\gamma_{+}-\gamma_{-}}\Big(\frac{1}{\beta\gamma_{\pm}}+\frac{i}{2}\Big).
\ee
\allowdisplaybreaks
The derivations of \Eqs{app_xleftaction}--(\ref{app_prightaction}) are as follows. Let us start with the actions of coordinate operator, $\hat x_{\B}$, in the Wigner representation [cf.\ \Eq{doublerho} with \Eq{aadg}],
\begin{align*}
	\hat x_{\B}^{\gler}W(x_{\B},p_{\B})
	=     & \Big(x_{\B}\pm\frac{i}{2}\frac{\partial}{\partial p_{\B}}\Big)W(x_{\B},p_{\B}),
	\nl = & \left(\frac{\beta\w_{\B}}{2\pi}\right)^{\frac{1}{2}}\sum_{n_{1},n_{2}}\frac{s_{n_{1},n_{2}}}{\sqrt{n_1!n_2!}}\rho_{n_{1},n_{2}}\Big(x_{\B}\pm\frac{i}{2}\frac{\partial}{\partial p_{\B}}\Big)
	\nl   & \times e^{-\frac{\beta\w_{\B}}{4}(x_{\B}^2+p_{\B}^2)}\Psi_{n_{1},n_{2}}
	\nl = &
	\left(\frac{\beta\w_{\B}}{2\pi}\right)^{\frac{1}{2}}\sum_{n_{1},n_{2}}\frac{s_{n_{1},n_{2}}}{\sqrt{n_1!n_2!}}\rho_{n_{1},n_{2}}e^{-\frac{\beta\w_{\B}}{4}(x_{\B}^2+p_{\B}^2)}
	\nl   & \times \Big(\frac{ a_1+ a_1^{\dg}}{\sqrt{\beta\w_{\B}}}\mp\frac{i\sqrt{\beta\w_{\B}}}{2} a_2^{\dg}\Big)\Psi_{n_{1},n_{2}}.
\end{align*}
Apply then \Eq{ccdg} and the FP algebra,\cite{Ris89, Din17024104} resulting in
\begin{align*}
	\hat x_{\B}^{\gler}W = & \left(\frac{\beta\w_{\B}}{2\pi}\right)^{\frac{1}{2}}\sum_{n_{1},n_{2}}\frac{s_{n_{1},n_{2}}}{\sqrt{n_1!n_2!}}\rho_{n_{1},n_{2}}e^{-\frac{\beta\w_{\B}}{4}(x_{\B}^2+p_{\B}^2)}
	\nl                    & \times \frac{1}{\sqrt{\beta\w_{\B}}}
	\Big[-r_2 c_1^{-}+r_1 c_2^{-}+(r_2\mp ir_1\beta\w_{\B}/2)c_1^{+}
	\nl                    & \ \ \ \ \ \ \ \ \ \ \ \
	+(r_1\mp ir_2\beta\w_{\B}/2)c_2^{+}\Big]\Psi_{n_{1},n_{2}}
	\nl =                  &
	\left(\frac{\beta\w_{\B}}{2\pi}\right)^{\frac{1}{2}}\sum_{n_{1},n_{2}}\frac{s_{n_{1},n_{2}}}{\sqrt{n_1!n_2!}}\rho_{n_{1},n_{2}}e^{-\frac{\beta\w_{\B}}{4}(x_{\B}^2+p_{\B}^2)}
	\nl                    & \times \frac{1}{\sqrt{\beta\w_{\B}}} \Big[-r_2 \sqrt{n_1}\Psi_{n_{1}-1,n_{2}}+r_1 \sqrt{n_2}\Psi_{n_{1},n_{2}-1}
	\nl                    & \ \ \ \ \ \ \ \ \ \ \ \
	+(r_2\mp ir_1\beta\w_{\B}/2)\sqrt{n_1+1}\Psi_{n_{1}+1,n_{2}}
	\nl                    & \ \ \ \ \ \ \ \ \ \ \ \
	+(r_1\mp ir_2\beta\w_{\B}/2)\sqrt{n_2+1}\Psi_{n_{1},n_{2}+1}\Big]
	\nl =                  &
	\left(\frac{\beta\w_{\B}}{2\pi}\right)^{\frac{1}{2}}e^{-\frac{\beta\w_{\B}}{4}(x_{\B}^2+p_{\B}^2)}\Big[(1)+(2)+(3)+(4)\Big],
\end{align*}
where
\begin{align*}
	(1) &
	=\sum_{n_{1},n_{2}}\frac{s_{n_{1}+1,n_{2}}}{\sqrt{(n_1+1)!n_2!}}\rho_{n_{1}+1,n_{2}}
	\frac{-r_2}{\sqrt{\beta\w_{\B}}} \sqrt{n_1+1}\Psi_{n_{1},n_{2}}
	\nl & =\sum_{n_{1},n_{2}}\frac{s_{n_{1},n_{2}}}{\sqrt{n_1!n_2!}}\rho_{n_{1}+1,n_{2}}\Psi_{n_{1},n_{2}}.
	\nl
	(2) &
	=\sum_{n_{1},n_{2}}\frac{s_{n_{1},n_{2}+1}}{\sqrt{n_1!(n_2+1)!}}\rho_{n_{1},n_{2}+1}
	\frac{r_1}{\sqrt{\beta\w_{\B}}} \sqrt{n_2+1}\Psi_{n_{1},n_{2}}
	\nl & =\sum_{n_{1},n_{2}}\frac{s_{n_{1},n_{2}}}{\sqrt{n_1!n_2!}}\rho_{n_{1},n_{2}+1}\Psi_{n_{1},n_{2}}.
	\nl
	(3) &
	=\sum_{n_{1},n_{2}}\frac{\sqrt{n_1}s_{n_{1}-1, n_{2}}}{\sqrt{(n_1-1)!n_2!}}\rho_{n_{1}-1,n_{2}}
	\Big(\frac{r_2\mp ir_1\beta\w_{\B}/2}{\sqrt{\beta\w_{\B}}}\Big)\Psi_{n_{1},n_{2}}
	\nl & =\sum_{n_{1},n_{2}}\frac{s_{n_{1}, n_{2}}}{\sqrt{n_1!n_2!}}\rho_{n_{1}-1,n_{2}}\Big(\frac{-r_2^2}{\beta\w_{\B}}\pm \frac{ir_1r_2 }{2}\Big)n_1\Psi_{n_{1},n_{2}}
	\nl &
	=\sum_{n_{1},n_{2}}\frac{s_{n_{1}, n_{2}}}{\sqrt{n_1!n_2!}}\rho_{n_{1}-1,n_{2}}(\eta_{+}/\bar\eta^{\ast}_{+})n_1\Psi_{n_{1},n_{2}}.
	\nl
	(4) &
	=\sum_{n_{1},n_{2}}\frac{\sqrt{n_2}s_{n_{1},n_{2}-1}}{\sqrt{n_1!(n_2-1)!}}\rho_{n_{1},n_{2}-1}
	\Big(\frac{r_1\mp ir_2\beta\w_{\B}/2}{\sqrt{\beta\w_{\B}}}\Big)\Psi_{n_{1},n_{2}}
	\nl & =\sum_{n_{1},n_{2}}\frac{s_{n_{1},n_{2}}}{\sqrt{n_1!n_2!}}\rho_{n_{1},n_{2}-1}\Big(\frac{r_1^2}{\beta\w_{\B}}\mp \frac{ir_1r_2 }{2}\Big)n_2\Psi_{n_{1},n_{2}}
	\nl &
	=\sum_{n_{1},n_{2}}\frac{s_{n_{1},n_{2}}}{\sqrt{n_1!n_2!}}\rho_{n_{1},n_{2}-1}(\eta_{-}/\bar\eta^{\ast}_{-})n_2\Psi_{n_{1},n_{2}}.
\end{align*}
In the last steps of above derivations, \Eqs{etaHT} and (\ref{r12}) are used.
We then finish deriving  \Eqs{app_xleftaction} and (\ref{app_xrightaction}) by adding the four terms together.
Similarly, for actions of $p_{\B}^{\gler}$ on the coefficient function $\rho_{n_{1},n_{2}}$, we obtain \Eqs{app_pleftaction} and (\ref{app_prightaction}).

Turn to core--system EOM in \Eq{hdeom} with \Eq{L-core}. 
By using \Eqs{app_xleftaction}-(\ref{app_prightaction}), we obtain
\begin{align} \label{final}
	\dot \rho_{{\bf n}; {\ti{\bf n}}} = &
	- \Big[i {\cal L}_{\tS} - 2 (n_{1}-n_{2}) \w_{\B} r_{1}r_{2} + \gamma_{\,\ti{\bf  n}}
	\nl                                &
	\quad \ + i \sum_{j=1}^{2} (2n_{j} + 1) ({\ti \lambda} C_{j} + \alpha_2 {\mathcal{C}_{j}}) + i \alpha_0 {\mathcal{A}} \Big] \rho_{{\bf n};{\ti{\bf n}}}
	\nl                                &
	- \sum_{j=1}^{2} \Big[ (-1)^{j} n_{j} \w_{\B} r_{1}r_{2} (1+{\gamma_{\bar j}}/{\gamma_{j}})
	\nl                                &
	\quad \quad \ \ + 2 i n_{j} ({\ti \lambda} C_{j} + \alpha_2 {\mathcal{C}}_{j}) \Big]\rho_{{\bf n}_{\,j\,{\bar j}}^{\minus\plus};{\ti{\bf n}}}
	\nl                                &
	- i \alpha_1 \sum_{j=1}^{2} \Big( {\mathcal{A}} \rho_{{\bf n}_{j}^{+};{\ti{\bf n}}} + n_{j} {\mathcal{C}}_{j}\rho_{{\bf n}_j^{-};{\ti{\bf n}}}\Big) - i \sum_{j=1}^{2} \sum_{k=1}^{N_K} n_{j} C_{j} \rho_{{\bf n}_{j}^{-};{\ti{\bf n}}_{k}^+}
	\nl                                & - i \sum_{j=1}^{2} \sum_{k=1}^{N_K} {\bm \ti n}_{k} \Big(C_{k} \rho_{{\bf n}_{j}^{+}; {\ti{\bf  n}}_{k}^{-}} + n_j \ti{B}_{j k} \rho_{{\bf n}_{j}^{-}; {\ti{\bf  n}}_{k}^{-}} \Big)
	\nl                                &
	- i \sum_{j=1}^{2} \sum_{j'=1}^{2} n_{j}(n_{j'} - \delta_{j j'}) ({\ti \lambda} B_{j j'} + \alpha_2 {\mathcal B}_{j j'})\rho_{{\bf n}_{j j'}^{-\ -};{\ti{\bf n}}}
	\nl                                &
	- i \alpha_2 \sum_{j=1}^{2} \sum_{j'=1}^{2} {\mathcal{A}} \rho_{{\bf n}_{j j'}^{+\ +};{\ti{\bf n}}},
\end{align}
with $\gamma_{1/2} = \gamma_{\pm}$ and $\ti \lambda$ being given in \Eq{ti_lambda},
\begin{equation}
	\begin{split}
		B_{j j'} &= \eta_{j} \eta_{j'} - {\bar\eta}_{j}^{\ast} {\bar\eta}_{j'}^{\ast} , \ \ \,
		{C}_{j} = \eta_{j} - {\bar\eta}_{j}^{\ast}, \\
		\ti{B}_{j k} &= \eta_{j} \ti \eta_{k} - \bar {\eta}_{j}^{\ast} {\ti \eta}_{\bar k}^{\ast} ,\ \ \,
		\quad \, \ti{C}_{k} = \ti \eta_{k} - {\ti \eta}_{\bar k}^{\ast}.
	\end{split}
\end{equation}
The involved superoperators are similar to those of \Eq{calABC}, but with
\begin{equation}
	\begin{split}
		&{\mathcal A} \hat O = \hat Q_{\tS} \hat O - \hat O \hat Q_{\tS},\ \ \,
		{\mathcal C}_{j} \hat O = \eta_{j} \hat Q_{\tS} \hat O - {\bar\eta}_{j}^{\ast} \hat O \hat Q_{\tS}, \\
		&{\mathcal B}_{jj'} \hat O = \eta_{j} \eta_{j'} \hat Q_{\tS} \hat O - {\bar\eta}_{j}^{\ast} {\bar\eta}_{j'}^{\ast} \hat O \hat Q_{\tS}.
	\end{split}
\end{equation}
Note also that $\bar j = 2$ when $j = 1$, and vice versa.


\begin{thebibliography}{10}

\bibitem{Wei21}
U.~Weiss,
\newblock {\em Quantum Dissipative Systems},
\newblock World Scientific, Singapore, 2021,
\newblock 5$^{\rm th}$ ed.

\bibitem{Kle09}
H.~Kleinert,
\newblock {\em Path Integrals in Quantum Mechanics, Statistics, Polymer
  Physics, and Financial Markets},
\newblock World Scientific, Singapore, 5th edition, 2009.

\bibitem{Bre02}
H.~P. Breuer and F.~Petruccione,
\newblock {\em The Theory of Open Quantum Systems},
\newblock Oxford University Press, New York, 2002.

\bibitem{Yan05187}
Y.~J. Yan and R.~X. Xu, \newblock ``Quantum mechanics of dissipative systems,''
  Annu. Rev. Phys. Chem. {\bf 56}, 187 (2005).

\bibitem{Nit06}
A.~Nitzan,
\newblock {\em Chemical Dynamics in Condensed Phases: Relaxation, Transfer and
  Reactions in Condensed Molecular Systems},
\newblock Oxford University Press, New York, 2006.

\bibitem{Muk95}
S.~Mukamel,
\newblock {\em The Principles of Nonlinear Optical Spectroscopy},
\newblock Oxford University Press, New York, 1995.

\bibitem{Lou73}
W.~H. Louisell,
\newblock {\em Quantum Statistical Properties of Radiation},
\newblock Wiley, New York, 1973.

\bibitem{Haa7398}
F.~Haake, \newblock ``Statistical treatment of open systems by generalized
  master equations,'' in {\em Quantum Statistics in Optics and Solid State
  Physics: Springer Tracts in Modern Physics, Vol.~66}, edited by
  G.~{H\"{o}hler}, pages 98--168, Springer, Berlin, 1973.

\bibitem{Aka15056002}
Y.~Akamatsu, \newblock ``Heavy quark master equations in the Lindblad form at
  high temperatures,'' Phys. Rev. D {\bf 91}, 056002 (2015).

\bibitem{Che964565}
V.~Chernyak and S.~Mukamel, \newblock ``Collective coordinates for nuclear
  spectral densities in energy transfer and femtosecond spectroscopy of
  molecular aggregates,'' J. Chem. Phys. {\bf 105}, 4565 (1996).

\bibitem{Imr02}
I.~Imry,
\newblock {\em Introduction to Mesoscopic Physics},
\newblock Oxford university press, 2002.

\bibitem{Hau08}
H.~Haug and A.-P. Jauho,
\newblock {\em Quantum Kinetics in Transport and Optics of Semiconductors},
\newblock Springer-Verlag, Berlin, 2nd, substantially revised edition, 2008,
\newblock Springer Series in Solid-State Sciences 123.

\bibitem{Fey63118}
R.~P. Feynman and F.~L. \mbox{Vernon, Jr.}, \newblock ``The theory of a general
  quantum system interacting with a linear dissipative system,'' Ann. Phys.
  {\bf 24}, 118 (1963).

\bibitem{Tan906676}
Y.~Tanimura, \newblock ``Nonperturbative expansion method for a quantum system
  coupled to a harmonic-oscillator bath,'' Phys. Rev. A {\bf 41}, 6676 (1990).

\bibitem{Tan06082001}
Y.~Tanimura, \newblock ``Stochastic Liouville, Langevin, Fokker-Planck, and
  master equation approaches to quantum dissipative systems,'' J. Phys. Soc.
  Jpn. {\bf 75}, 082001 (2006).

\bibitem{Yan04216}
Y.~A. Yan, F.~Yang, Y.~Liu, and J.~S. Shao, \newblock ``Hierarchical approach
  based on stochastic decoupling to dissipative systems,'' Chem. Phys. Lett.
  {\bf 395}, 216 (2004).

\bibitem{Xu05041103}
R.~X. Xu, P.~Cui, X.~Q. Li, Y.~Mo, and Y.~J. Yan, \newblock ``Exact quantum
  master equation via the calculus on path integrals,'' J. Chem. Phys. {\bf
  122}, 041103 (2005).

\bibitem{Xu07031107}
R.~X. Xu and Y.~J. Yan, \newblock ``Dynamics of quantum dissipation systems
  interacting with bosonic canonical bath: Hierarchical equations of motion
  approach,'' Phys. Rev. E {\bf 75}, 031107 (2007).

\bibitem{Jin08234703}
J.~S. Jin, X.~Zheng, and Y.~J. Yan, \newblock ``Exact dynamics of dissipative
  electronic systems and quantum transport: Hierarchical equations of motion
  approach,'' J. Chem. Phys. {\bf 128}, 234703 (2008).

\bibitem{Ye16608}
L.~Z. Ye, X.~L. Wang, D.~Hou, R.~X. Xu, X.~Zheng, and Y.~J. Yan, \newblock
  ``HEOM-QUICK: A program for accurate, efficient and universal
  characterization of strongly correlated quantum impurity systems,'' WIREs
  Comp. Mol. Sci. {\bf 6}, 608 (2016).

\bibitem{Yan16110306}
Y.~J. Yan, J.~S. Jin, R.~X. Xu, and X.~Zheng, \newblock ``Dissipaton equation
  of motion approach to open quantum systems,'' Frontiers Phys. {\bf 11},
  110306 (2016).

\bibitem{Yan20204109}
Y.-M. Yan, T.~Xing, and Q.~Shi, \newblock ``A new method to improve the
  numerical stability of the hierarchical equations of motion for discrete
  harmonic oscillator modes,'' J. Chem. Phys. {\bf 153}, 204109 (2020).

\bibitem{Tan914131}
Y.~Tanimura and P.~G. Wolynes, \newblock ``Quantum and classical Fokker-Planck
  equations for a Guassian-Markovian noise bath,'' Phys. Rev. A {\bf 43}, 4131
  (1991).

\bibitem{Tan943049}
Y.~Tanimura and S.~Mukamel, \newblock ``Multistate quantum Fokker-Planck
  approach to nonadiabatic wave packet dynamics in pump-probe spectroscopy,''
  J. Chem. Phys. {\bf 101}, 3049 (1994).

\bibitem{Tan971779}
Y.~Tanimura and Y.~Maruyama, \newblock ``Gaussian-Markovian quantum
  Fokker-Planck approach to nonlinear spectroscopy of a displaced Morse
  potentials system: dissociation, predissociation, and optical Stark
  effects,'' J. Chem. Phys. {\bf 107}, 1779 (1997).

\bibitem{Ike17014102}
T.~Ikeda and Y.~Tanimura, \newblock ``Probing photoisomerization processes by
  means of multi-dimensional electronic spectroscopy: The multi-state quantum
  hierarchical Fokker-Planck equation approach,'' J. Chem. Phys. {\bf 147},
  014102 (2017).

\bibitem{Ike192517}
T.~Ikeda and Y.~Tanimura, \newblock ``Low-Temperature Quantum Fokker–Planck
  and Smoluchowski Equations and Their Extension to Multistate Systems,'' J.
  Chem. Theory Comput. {\bf 15}, 2517 (2019).

\bibitem{Tan15144110}
Y.~Tanimura, \newblock ``Real-time and imaginary-time quantum hierarchal
  Fokker-Planck equations,'' J. Chem. Phys. {\bf 142}, 144110 (2015).

\bibitem{Li22064107}
T.-C. Li, Y.-M. Yan, and Q.~Shi, \newblock ``A low-temperature quantum
  Fokker-Planck equation that improves the numerical stability of the
  hierarchical equations of motion for the Brownian oscillator spectral
  density,'' J. Chem. Phys. {\bf 156}, 064107 (2022).

\bibitem{Ike22104104}
T.~Ikeda and A.~Nakayama, \newblock ``Collective bath coordinate mapping of
  “hierarchy” in hierarchical equations of motion,'' J. Chem. Phys. {\bf
  156}, 104104 (2022).

\bibitem{Vio02886}
D.~Vion, A.~Aassime, A.~Cottet, P.~Joyez, H.~Pothier, C.~Urbina, D.~Esteve, and
  M.~H. Devoret, \newblock ``Manipulating the quantum state of an electrical
  circuit,'' Science {\bf 296}, 886 (2002).

\bibitem{Mak04178301}
Y.~Makhlin and A.~Shnirman, \newblock ``Dephasing of solid-state qubits at
  optimal points,'' Phys. Rev. Lett. {\bf 92}, 178301 (2004).

\bibitem{Mul04237401}
E.~A. Muljarov and R.~Zimmermann, \newblock ``Dephasing in quantum dots:
  Quadratic coupling to acoustic phonons,'' Phys. Rev. Lett. {\bf 93}, 237401
  (2004).

\bibitem{Ber05257002}
P.~Bertet, I.~Chiorescu, G.~Burkard, K.~Semba, C.~J. P.~M. Harmans, D.~P.
  DiVincenzo, and J.~E. Mooij, \newblock ``Dephasing of a superconducting qubit
  induced by photon noise,'' Phys. Rev. Lett. {\bf 95}, 257002 (2005).

\bibitem{Yan865908}
Y.~J. Yan and S.~Mukamel, \newblock ``Eigenstate-free, Green function:
  Calculation of molecular absorption and fluorescence line shapes,'' J. Chem.
  Phys. {\bf 85}, 5908 (1986).

\bibitem{Pen07114302}
Q.~Peng, Y.~P. Yi, Z.~G. Shuai, and J.~S. Shao, \newblock ``Excited state
  radiationless decay process with Duschinsky rotation effect: formalism and
  implementation,'' J. Chem. Phys. {\bf 126}, 114302 (2007).

\bibitem{Wan0710369}
H.~Wang and M.~Thoss, \newblock ``Quantum Dynamical Simulation of
  Electron-Transfer Reactions in an Anharmonic Environment,'' J. Phys. Chem. A
  {\bf 111}, 10369 (2007).

\bibitem{Zha121075}
Y.~Zhao and W.~Z. Liang, \newblock ``Charge transfer in organic molecules for
  solar cells: theoretical perspective,'' Chem. Soc. Rev {\bf 41}, 1075 (2012).

\bibitem{Cho17074114}
V.~Choro$\check{\rm s}$ajev, T.~Mar$\check{\rm c}$iulionis, and
  D.~Abramavicius, \newblock ``Temporal dynamics of excitonic states with
  nonlinear electron-vibrational coupling,'' J. Chem. Phys. {\bf 147}, 074114
  (2017).

\bibitem{Wan21462}
Y.~Wang, Y.~Su, R.~X. Xu, X.~Zheng, and Y.~J. Yan, \newblock ``Marcus' electron
  transfer rate revisited via a generalized Rice--Ramsperger--Kassel--Marcus
  theory with nonlinear environments,'' Chin. J. Chem. Phys. {\bf 34}, 462
  (2021).

\bibitem{Xu17395}
R.~X. Xu, Y.~Liu, H.~D. Zhang, and Y.~J. Yan, \newblock ``Theory of quantum
  dissipation in a class of non-Gaussian environments,'' Chin. J. Chem. Phys.
  {\bf 30}, 395 (2017).

\bibitem{Xu18114103}
R.~X. Xu, Y.~Liu, H.~D. Zhang, and Y.~J. Yan, \newblock ``Theories of quantum
  dissipation and nonlinear coupling bath descriptors,'' J. Chem. Phys. {\bf
  148}, 114103 (2018).

\bibitem{Yan19074106}
Y.~A. Yan, \newblock ``Stochastic simulation of anharmonic dissipation. II.
  Harmonic bath potentials with quadratic couplings,'' J. Chem. Phys. {\bf
  150}, 074106 (2019).

\bibitem{Hsi18014104}
C.-Y. Hsieh and J.~Cao, \newblock ``A unified stochastic formulation of
  dissipative quantum dynamics. II. Beyond linear response of spin baths,'' J.
  Chem. Phys. {\bf 148}, 014104 (2018).

\bibitem{Hsi20125002}
J.~T. Hsiang and B.~L. Hu, \newblock ``Nonequilibrium nonlinear open quantum
  systems: Functional perturbative analysis of a weakly anharmonic
  oscillator,'' Phys. Rev. D {\bf 101}, 125002 (2020).

\bibitem{Hsi20125003}
J.~T. Hsiang and B.~L. Hu, \newblock ``Fluctuation-dissipation relation from
  the nonequilibrium dynamics of a nonlinear open quantum system,'' Phys. Rev.
  D {\bf 101}, 125003 (2020).

\bibitem{Yan14054105}
Y.~J. Yan, \newblock ``Theory of open quantum systems with bath of electrons
  and phonons and spins: Many-dissipaton density matrixes approach,'' J. Chem.
  Phys. {\bf 140}, 054105 (2014).

\bibitem{Xu151816}
R.~X. Xu, H.~D. Zhang, X.~Zheng, and Y.~J. Yan, \newblock ``Dissipaton equation
  of motion for system-and-bath interference dynamics,'' Sci. China Chem. {\bf
  58}, 1816 (2015),
\newblock Special Issue: Lemin Li Festschrift.

\bibitem{Zha18780}
H.~D. Zhang, R.~X. Xu, X.~Zheng, and Y.~J. Yan, \newblock ``Statistical
  quasi-particle theory for open quantum systems,'' Mol. Phys. {\bf 116}, 780
  (2018),
\newblock Special Issue, ``Molecular Physics in China''.

\bibitem{Wan20041102}
Y.~Wang, R.~X. Xu, and Y.~J. Yan, \newblock ``Entangled system-and-environment
  dynamics: Phase-space dissipaton theory,'' J. Chem. Phys. {\bf 152}, 041102
  (2020).

\bibitem{Wan22170901}
Y.~Wang and Y.~J. Yan, \newblock ``Quantum mechanics of open systems:
  Dissipaton theories,'' J. Chem. Phys. {\bf 157}, 170901 (2022).

\bibitem{Zha15024112}
H.~D. Zhang, R.~X. Xu, X.~Zheng, and Y.~J. Yan, \newblock ``Nonperturbative
  spin-boson and spin-spin dynamics and nonlinear Fano interferences: A unified
  dissipaton theory based study,'' J. Chem. Phys. {\bf 142}, 024112 (2015).

\bibitem{Che21244105}
Z.~H. Chen, Y.~Wang, R.~X. Xu, and Y.~J. Yan, \newblock ``Correlated
  vibration-solvent effects on the non-Condon exciton spectroscopy,'' J. Chem.
  Phys. {\bf 154}, 244105 (2021).

\bibitem{Che21174111}
Z.~H. Chen, Y.~Wang, R.~X. Xu, and Y.~J. Yan, \newblock ``Quantum dissipation
  with nonlinear environment couplings: Stochastic fields dressed dissipaton
  equation of motion approach,'' J. Chem. Phys. {\bf 155}, 174111 (2021).

\bibitem{Ris89}
H.~Risken,
\newblock {\em The Fokker-Planck Equation, Methods of Solution and
  Applications},
\newblock Springer-Verlag, Berlin, 2nd edition, 1989.

\bibitem{Xu09074107}
R.~X. Xu, B.~L. Tian, J.~Xu, and Y.~J. Yan, \newblock ``Exact dynamics of
  driven Brownian oscillators,'' J. Chem. Phys. {\bf 130}, 074107 (2009).

\bibitem{Din17024104}
J.~J. Ding, Y.~Wang, H.~D. Zhang, R.~X. Xu, X.~Zheng, and Y.~J. Yan, \newblock
  ``Fokker-Planck quantum master equation for mixed quantum-semiclassical
  dynamics,'' J. Chem. Phys. {\bf 146}, 024104 (2017).

\bibitem{Gon20214115}
H.~Gong, Y.~Wang, H.~D. Zhang, R.~X. Xu, X.~Zheng, and Y.~J. Yan, \newblock
  ``Thermodynamic free--energy spectrum theory for open quantum systems,'' J.
  Chem. Phys. {\bf 153}, 214115 (2020).

\bibitem{Tan14044114}
Y.~Tanimura, \newblock ``Reduced hierarchical equations of motion in real and
  imaginary time: Correlated initial states and thermodynamic quantities,'' J.
  Chem. Phys. {\bf 141}, 044114 (2014).

\bibitem{Gon22054109}
H.~Gong, Y.~Wang, X.~Zheng, R.~X. Xu, and Y.~J. Yan, \newblock ``Nonequilibrium
  work distributions in quantum impurity system--bath mixing processes,'' J.
  Chem. Phys. {\bf 157}, 054109 (2022).

\bibitem{Hu10101106}
J.~Hu, R.~X. Xu, and Y.~J. Yan, \newblock ``Pad\'{e} spectrum decomposition of
  Fermi function and Bose function,'' J. Chem. Phys. {\bf 133}, 101106 (2010).

\bibitem{Hu11244106}
J.~Hu, M.~Luo, F.~Jiang, R.~X. Xu, and Y.~J. Yan, \newblock ``Pad\'{e} spectrum
  decompositions of quantum distribution functions and optimal hierarchial
  equations of motion construction for quantum open systems,'' J. Chem. Phys.
  {\bf 134}, 244106 (2011).

\bibitem{Din12224103}
J.~J. Ding, R.~X. Xu, and Y.~J. Yan, \newblock ``Optimizing hierarchical
  equations of motion for quantum dissipation and quantifying quantum bath
  effects on quantum transfer mechanisms,'' J. Chem. Phys. {\bf 136}, 224103
  (2012).

\bibitem{Fan112145}
H.~Fan, H.~Yuan, and N.~Jiang, \newblock ``New identities about operator
  Hermite polynomials and their related integration formulas,'' Sci. China
  Phys. Mech. Astron. {\bf 54}, 2145 (2011).

\bibitem{Wan20}
Y.~Wang, \newblock ``Quantum Mechanics of Open Systems: Dissipaton Theory (in
  Chinese),'' PhD Thesis, University of Science and Technology of China
  (2020),
\newblock DOI:10.27517/d.cnki.gzkju.2020.000617; See also:
  http://home.ustc.edu.cn/$\sim$wy2010/thesis.pdf.

\bibitem{Din16204110}
J.~J. Ding, H.~D. Zhang, Y.~Wang, R.~X. Xu, X.~Zheng, and Y.~J. Yan, \newblock
  ``Minimum--exponents ansatz for molecular dynamics and quantum dissipation,''
  J. Chem. Phys. {\bf 145}, 204110 (2016).

\bibitem{Che22221102}
Z.~H. Chen, Y.~Wang, X.~Zheng, R.~X. Xu, and Y.~J. Yan, \newblock ``Universal
  time-domain Prony fitting decomposition for optimized hierarchical quantum
  master equations,'' J. Chem. Phys. {\bf 156}, 221102 (2022).

\bibitem{Wig32749}
E.~Wigner, \newblock ``On the quantum correction for thermodynamic
  equilibrium,'' Phys. Rev. {\bf 40}, 749 (1932).

\bibitem{Liu18245}
Y.~Liu, R.~X. Xu, H.~D. Zhang, and Y.~J. Yan, \newblock ``Dissipaton equation
  of motion theory versus Fokker-Planck quantum master equation,'' Chin. J.
  Chem. Phys. {\bf 31}, 245 (2018).

\bibitem{Cal83587}
A.~O. Caldeira and A.~J. Leggett, \newblock ``Path integral approach to quantum
  Brownian motion,'' Physica A {\bf 121}, 587 (1983).

\bibitem{Shi09084105}
Q.~Shi, L.~P. Chen, G.~J. Nan, R.~X. Xu, and Y.~J. Yan, \newblock ``Efficient
  hierarchical Liouville space propagator to quantum dissipative dynamics,'' J.
  Chem. Phys. {\bf 130}, 084105 (2009).

\bibitem{Jar11329}
C.~Jarzynski, \newblock ``Equalities and inequalities: Irreversibility and the
  Second Law of Thermodynamics at the nanoscale,'' Ann. Rev. Cond. Matter Phys.
  {\bf 2}, 329 (2011).

\bibitem{Cro992721}
G.~E. Crooks, \newblock ``Entropy production fluctuation theorem and the
  nonequilibrium work relation for free energy differences,'' Phys. Rev. E {\bf
  60}, 2721 (1999).

\bibitem{Tal07F569}
P.~Talkner and P.~H\"{a}nggi, \newblock ``The Tasaki--Crooks quantum
  fluctuation theorem,'' J. Phys. A: Math. Theor. {\bf 40}, F569 (2007).

\bibitem{Tal07050102}
P.~Talkner, E.~Lutz, and P.~H\"{a}nggi, \newblock ``Fluctuation theorems: Work
  is not an observable,'' Phys. Rev. E {\bf 75}, 050102 (2007).

\bibitem{Sak21033001}
S.~Sakamoto and Y.~Tanimura, \newblock ``Open quantum dynamics theory for
  non-equilibrium work: Hierarchical equations of motion approach,'' J. Phys.
  Soc. Jpn. {\bf 90}, 033001 (2021).

\bibitem{Esp091665}
M.~Esposito, U.~Harbola, and S.~Mukamel, \newblock ``Nonequilibrium
  fluctuations, fluctuation theorems, and counting statistics in quantum
  systems,'' Rev. Mod. Phys. {\bf 81}, 1665 (2009).

\bibitem{Du20034102}
P.~L. Du, Y.~Wang, R.~X. Xu, H.~D. Zhang, and Y.~J. Yan, \newblock
  ``System-bath entanglement theorem with Gaussian environments,'' J. Chem.
  Phys. {\bf 152}, 034102 (2020).

\bibitem{Wan22044102}
Y.~Wang, Z.~H. Chen, R.~X. Xu, X.~Zheng, and Y.~J. Yan, \newblock ``A
  statistical quasi--particles thermofield theory with Gaussian environments:
  System--bath entanglement theorem for nonequilibrium correlation functions,''
  J. Chem. Phys. {\bf 157}, 044012 (2022).

\bibitem{Scu97}
M.~O. Scully and M.~S. Zubairy,
\newblock {\em Quantum Optics},
\newblock Cambridge University Press, Cambridge, 1997.

\end{thebibliography}

\end{document}